\documentclass{article}

\usepackage{arxiv}

\usepackage[utf8]{inputenc} 
\usepackage[T1]{fontenc}    
\usepackage{hyperref}       
\usepackage{url}            
\usepackage{booktabs}       
\usepackage{amsfonts}       
\usepackage{nicefrac}       
\usepackage{microtype}      
\usepackage{lipsum}
\usepackage{graphicx}
\usepackage{algorithm}
\usepackage{algorithmic}
\usepackage{bm}
\usepackage{array}
\usepackage{tabularx}
\usepackage{mwe}
\graphicspath{ {./images/} }

\title{Stochastic Step-wise Feature Selection for  Exponential Random Graph Models (ERGMs)}

\author{
 Helal El-Zaatari \\
  Department of Biostatistics\\
  University of North Carolina\\
  Chapel Hill, NC 27514 \\
  \texttt{helal@live.unc.edu} \\
   \And
 Fei Yu \\
  Helath Sciences Library\\
  University of North Carolina\\
  Chapel Hill, NC 27514\\
  \texttt{feifei@unc.edu} \\
  \And
 Michael R.Kosorok \\
  Department of Biostatistics\\
  University of North Carolina\\
  Chapel Hill, NC 27514 \\
  \texttt{kosorok@bios.unc.edu} \\
}

\begin{document}
\maketitle
\begin{abstract}
Statistical analysis of social networks provides valuable insights into complex network interactions across various scientific disciplines. However, accurate modeling of networks remains challenging due to the heavy computational burden and the need to account for observed network dependencies. Exponential Random Graph Models (ERGMs) have emerged as a promising technique used in social network modeling to capture network dependencies by incorporating endogenous variables. Nevertheless, using ERGMs poses multiple challenges, including the occurrence of ERGM degeneracy, which generates unrealistic and meaningless network structures. To address these challenges and enhance the modeling of collaboration networks, we propose and test a novel approach that focuses on endogenous variable selection within ERGMs. Our method aims to overcome the computational burden and improve the accommodation of observed network dependencies, thereby facilitating more accurate and meaningful interpretations of network phenomena in various scientific fields. We conduct empirical testing and rigorous analysis to contribute to the advancement of statistical techniques and offer practical insights for network analysis. 
\end{abstract}


\section{Introduction}

Statistical analysis of social networks plays a pivotal role in various scientific disciplines, offering valuable insights into complex network interactions. Accurate modeling is particularly crucial when working with moderately sized networks, typically comprising a few thousand nodes, as it enables the explanation, analysis, replication, and prediction of network phenomena observed in nature. In the field of health sciences, social network analysis contributes to reducing health disparities \cite{okamoto2015scientific} and fostering collaboration and research efficiency, leading to scientific innovations and discoveries\cite{bennett2018collaboration}.By uncovering patterns in collaboration networks, network analysis facilitates the prediction of future connections among individuals or organizations, which holds significant value for multiple stakeholders including health policy researchers, administrators, and research sponsors \cite{yu2020bibliometrics}  \cite{provan2005network} \cite{luke2013network}. \newline 



The advancement in computational power of personal computers in the 21st century has empowered researchers to conduct sophisticated statistical modeling without relying on supercomputers \cite{nordhaus2001progress}. One powerful technique widely used in social network research is Exponential Random Graph Models (ERGMs). ERGMs effectively capture network dependencies by incorporating endogenous variables. However, a challenge arises when the chosen endogenous variables fail to accurately capture the structures present in the observed network, resulting in a phenomenon known as ERGM degeneracy \cite{li2015degeneracy}. ERGM degeneracy leads to the generation of unrealistic and meaningless network structures, where complete graphs are proposed with all nodes connected to one another \cite{krivitsky2017using} \cite{li2015degeneracy} \cite{bang2018basic}. \newline

Addressing the weakness of ERGMs presents multiple challenges that require careful consideration. First, the dependency among observations in network data renders statistical models assuming independence invalid \cite{kolaczyk2014statistical}. This dependency issue shares similarities with the challenges encountered in analyzing longitudinal studies, where dependent observations arise from repeated measurements. Secondly, accurately modeling the complex dependency among observations is both crucial and challenging for understanding the network structure. While stochastic block models treat this dependency as a nuisance parameter, ERGMs consider it as a phenomenon to be explicitly modeled and quantified through endogenous variables. However, modeling this dependency is a demanding task, even for experienced statisticians, let alone researchers with limited knowledge of advanced statistical modeling techniques. Moreover, ERGMs encompass various types and variations, which introduce challenges in selecting appropriate endogenous variables and screening for degeneracy. The existing literature reports at least five distinct types of ERGMs: the standard ERGM  \cite{uddin2013study} \cite{zappa2011interplay}, Bayesian ERGM \cite{caimo2017bayesian}, Temporal ERGM \cite{azondekon2018modeling}, Separable Temporal ERGM \cite{ho2021fostering} \cite{broekel2018disentangling}, and Multi-level ERGM \cite{wang2013exponential} \cite{mcglashan2019collaboration}. All these extensions of ERGMs require the incorporation of endogenous variables. However, selecting the suitable endogenous variable for a given observed network poses a significant challenge. There are thousands of potential endogenous variables to choose from \cite{hunter2008ergm} which leaves many researchers overwhelmed. Additionally, choosing the appropriate endogenous variable can be a perilous task as even the addition of a single unnecessary endogenous variable can lead to ERGM degeneracy. Unfortunately, researchers currently lack algorithms or tools to assist in the selection of endogenous variables. \newline 

Therefore, the primary objective of this study is to propose and test a novel approach that specifically addresses the significant challenges associated with ERGMs in modeling collaboration networks. By addressing the issues of dependency among observations, complex dependency modeling, endogenous variable selection, and degeneracy screening, we aim to provide a comprehensive solution that enhances the effectiveness and reliability of ERGM modeling for collaboration networks. Through empirical testing and rigorous analysis, we seek to contribute to the advancement of statistical techniques and facilitate more accurate and meaningful interpretations of network phenomena in various scientific disciplines.\newline 

The structure of this paper is outlined as follows. In Section 2, we provide the mathematical foundation upon which the three algorithms of this study are built. We begin by rigorously defining ERGMs through random graphs, explicitly characterizing the probability measure that generates this class of models. Additionally, we justify the use of edges, 2-stars, and triangle counts as measures to capture the structure of observed networks by introducing the concept of homomorphism densities \cite{chatterjee2013estimating}. Section 3 delves into the details of the endogenous variable selection procedure, which consists of three main parts. First, we discuss the process of obtaining the initial set of endogenous variables. Second, we present a novel step-wise variable selection procedure that accommodates for fluctuating likelihood estimates caused by the ERGMs’ intractable normalizing constant. Third, we describe the degeneracy screening procedure, which is based on the concept of homomorphism densities. In Section 4, we apply the three algorithms proposed in this paper to eleven real-life un-directed binary networks. We present numerical results, including the number of selected endogenous variables, the count of potential pairwise ERGMs, average counts of edges, 2-stars, and triangles, as well as the number of screened degenerate ERGMs. Finally, the last section provides a discussion on future research directions and potential extensions of the methods presented in this study.

\section{Mathematical Specification of ERGMs}

A network or graph $G$ consists of nodes and edges denoted by $G=(V,E)$ respectively. The nodes are assumed to be finite with $V= \{1, \dots, N \}$. The edges represent ties between two different nodes $i,j$. Modeling networks is centered around the edges $E$ of a graph. The outcome of interest $Y_{i,j}$ is defined for two separate nodes $i \in V$ and $j \in V$. Depending on the type of network the outcome $Y_{i,j}$ can take on binary, discrete or real valued numbers. For example, consider a binary outcome where $Y_{i,j}=1$ indicates an edge between nodes $i$ and $j$ and $Y_{i,j}=0$ indicates no edge. Nodes $i \in V$ are allowed to possess a collection of attributes in euclidean space. 

\subsection{Probabilistic Description of ERGMs}

 A model $\mathcal{M}$ for a network is a collection that is characterized by the ensemble of all possible graphs, denoted by $\mathcal{G}$, the parameter space $\Theta$ and a probability distribution $\mathbb{P}_{\theta}$ on $\mathcal{G}$.

\begin{equation}
    \mathcal{M} = \{ P_{\theta}(G) |\hspace{1em} G\in \mathcal{G}, \hspace{1em} \theta \in \Theta \}
\end{equation}

The choice of probability distribution will determine the complexity and richness of the network model. Inspired by generalized linear models, exponential random graphs model the probability of a tie formation $Y_{i,j} = 1$ given the nodal attributes $X$. For a binary outcome, exponential random graphs can be viewed as the logistic regression analogue for network data where the probability of a tie is allowed to depend on the presence of other ties in the network \cite{yousefi2014social} \cite{van2019introduction}. Similar analogies can be made between ERGMs with discrete and continuous valued ties with their generalized linear model analogue such as Poisson regression and Gamma regression respectively. 

\subsubsection{Defining the Probability Distribution for ERGMs}

Consider the space $\mathcal{G}_N$ of all graphs possessing $N \in \mathbb{N}$ vertices. We define the following probability distribution on this space. 

\begin{equation}
    p_N(Y) = \frac{exp\{ H(Y) \} }{\psi}
\end{equation}

$Y \in \mathcal{G}_N$, $\psi$ is the normalizing constant and $H(Y)$ is a Hamiltonian function used to weight the above probability measure. $H(Y)$ describes the total energy of a system of objects \cite{bhamidi2008mixing}. The Hamiltonian function for graphs is defined below.

\begin{equation}
    H(Y) = \sum_{i} \theta_i T_i(Y) 
\end{equation}

$T_i(Y)$ describes the different structures present in the graph. Now consider $s$ graphs $Y_1, \dots, Y_s$ where a graph $Y_i$ has $|V_i|$ labeled vertices. One can use the above definition of a probability distribution and apply it to exponential random graphs. Denote the observed network by $G$ and consider a graph $Y \in \mathcal{G}_N$. We define the following count statistic.

\begin{equation}
    N_G(Y) = \sum_{\boldsymbol{v}_m \in \{ 1,2, \dots, N  \}^m } I \{ H_Y( \boldsymbol{v}_m )  \simeq G \}
\end{equation}

 Here, $\boldsymbol{v}_m$ represents a set $m$ distinct vertices from $\{1,2,\dots, m \}$. The set $ \{ 1,2, \dots, N  \}^m $ denotes the distinct m-tuples. $H_Y(\boldsymbol{v}_m)$ denotes the sub-graph of $Y$ induced by $\boldsymbol{v}_m$. It is the sub-graph that is formed from the subset of vertices $\boldsymbol{v}_m$ of the graph and all the edges in the graph $Y$ that connect to the vertices found in $\boldsymbol{v}_m$. Multiplying the count statistic $N_G(Y)$ by coefficients $\theta_i$ and by defining $T(Y) = (s(Y), g(Y,X) )^T$, where $X$ represents the nodal attributes, one obtains the usual definition of an exponential random graph model \cite{bhamidi2008mixing}. 

\vspace{1em}

Using the above mathematically equivalent formulation of an exponential random graph model (ERGM). One can construct the following probability distribution for ERGMs in the space $\mathcal{G}_n$. 

\begin{equation}
    p_n(Y) = \frac{1}{\psi(\boldsymbol{\theta})} exp \{  \sum_i  \theta_i  \frac{N_{G}(Y_i)}{n^{|V_i| - 2}}  \} = \frac{h(y)\exp{\eta(\boldsymbol{\theta})^T g(y)}}{\psi} 
\end{equation}

Note that $\boldsymbol{\theta} = (\theta_1, \dots, \theta_s)$. The above probability distribution is also a Gibbs measure and the Hamiltonian for ERGMS is given by
$ H(Y) = \sum_i  \theta_i  \frac{N_{G}(Y_i)}{n^{|V_i| - 2}}$. The Hamiltonian plays a central role in ERGM   goodness of fit  as it's analytic properties, such as smoothness, influence the probability of producing degenerate network simulations from a candidate ERGM \cite{bhamidi2008mixing}.

\subsection{Homomorphism Densities}

Network motifs are small graphs, typically not exceeding 6 nodes, that are statistically significant patterns found in a larger network \cite{masoudi2012building}. From these network motif counts one can construct  homomorphism densities for an observed network and candidate ERGM. Homomorphisms are maps that preserve the algebraic structure found in a group \cite{dummit2004abstract}. Formally, for some group operation $\bigoplus $, the map $f: A \longrightarrow B$ is a homomorphism when it satisfies the relation below.

\begin{equation}
    f(x \bigoplus_A y) = f(x) \bigoplus_B f(y)  \hspace{1em} x,y \in A
\end{equation}

For finite un-directed graphs, $ G= (V(G), E(G))$ and $H = (V(H), E(H) )$, a homomorphism $f: G \longrightarrow H$ between the two graphs implies that for any two vertices $v_1$ and $v_2$ present in $E(G)$ we automatically have that $f(v_1)$ and $f(v_2)$ belong in $E(H)$ \cite{hahn1997graph}.  One can thus count the number of homomorphisms for a given network motif $H$ into an observed network $G$. Practically this counts the number of edge-preserving maps between vertex sets $V(H)$ and $V(G)$. Then after counting the number of maps, we can define the homomorphism density.

\begin{equation}
    t(H,G) = \frac{|hom(H,G)|}{|V(G)|^{|V(H)|}}
\end{equation}

This density represents the probability that any arbitrary mapping $t: V(H) \longrightarrow V(G)$ is edge preserving \cite{he2015glmle} \cite{chatterjee2013estimating}.

\section{Feature Selection for ERGMs}

An ERGM can be regarded as the logistic regression analogue for network data. 

\begin{equation}
    P_{\theta}(Y_{i,j}=1 | X) = \psi(\boldsymbol{\theta}_1 , \boldsymbol{\theta}_2) exp \{ \boldsymbol{\theta}_1^T \boldsymbol{s(y)}+ \boldsymbol{\theta}_2^T \boldsymbol{g(y,x)} \}
\end{equation}

$Y_{i,j}$ denotes the binary dependent variable of a tie formation between two distinct nodes denoted by $i$ and $j$. Nodes are allowed to posses attributes which are denoted via $X$. There are two types of independent variables in an ERGM. The first type are called endogenous variables and are a strict function of the existing ties $Y$. Endogenous variables are therefore denoted by $\boldsymbol{s(y)}$ with their associated regression coefficients $\boldsymbol{\theta_1}$. The second type of independent variables are called exogenous variables. These variables are a function of both nodal attributes and existing ties, although in most applications they are a strict function of the attributes $X$. Exogenous variables are denoted by $\boldsymbol{g(y,x)}$ with associated regression coefficients $\boldsymbol{\theta_2}$. The computationally expensive normalizing constant is given by $\psi$. \newline 

There exists hundreds of pre-defined endogenous variables one can include in an ERGM as well as customized user defined endogenous variables which model different network structures \cite{hunter2013ergm}. Unless one possess strong knowledge on the observed network's topology, the task of choosing the appropriate endogenous variables is very challenging. In this paper, thirteen commonly used endogenous variables, taken from the ERGM R package, populate the initial set of endogenous variables \cite{handcock2015package}. Namely: kstar, degree-wise shared partners (dsp), non-edgewise shared partners (nsp), edgewise shared partners (esp), triangle, isolates, sociality, degree cross product, degree popularity , geometrically weighted edgewise shared partners (gwesp), geometrically weighted non-edgewise shared partners (gwnsp), geometrically weighted dyad-wise shared partners (gwdsp) and geometrically weighted degree. These endogenous variables are specifically for binary un-directed networks. \newline 

In order to perform feature selection we need an initial set of endogenous variables. The main challenge of obtaining set, is that there are infinitely many endogenous variables to choose from. For example, the dyad-wise shared partners (dsp) is a network statistic equal to the number of dyads with  $k$ shared partners \cite{hunter2008ergm}. This endogenous variable requires an input $k \in \mathbb{N}$. However, one can not consider infinitely many inputs across the whole natural number line.  We propose a systematic way of obtaining such a set. The idea is to provide an informed upper bound for endogenous variables like dsp that require a natural number as an input.

\begin{figure}[ht]
    \centering
\includegraphics[width=14cm]{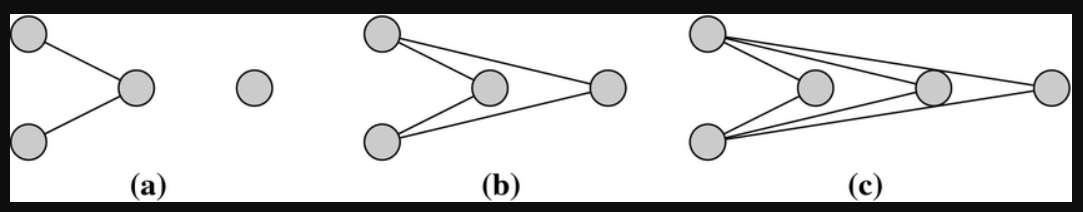}
    \caption{ Illustration of the dyadwise shared partner endogenous variable with $k=1$, $k=2$ and $k=3$ respectively. Source:  \cite{van2019introduction}. }
\end{figure}

After obtaining an initial set of endogenous variables, we apply a novel step-wise feature selection method in order to reduce the size of this initial set. This method is inspired by the classical statistical method of forward selection \cite{effroymson1960multiple}. However, several modifications are needed to accommodate for the intractable normalizing constant present in an ERGM's likelihood function. First, we only consider ERGMs with two predictors. The two predictors are the intercept and an element from our endogenous variable set. Then, we compare their Akaike Information Criterion (AIC) with the null ERGM possessing just the intercept. Second, we only exclude an endogenous variable from our model space if it consistently results in a large increase in AIC with respect to the null ERGM. This is quantified via the relative difference in AIC between the two nested models. Finally, we screen out degenerate ERGMs, which are characterized by an astronomical number of edges, based on the average count of edges generated by the ERGM. \newline

\subsection{Bounding endogenous variable that take on natural numbers}

Endogenous variables such as kstar, degree-wise shared partners (dsp), edge-wise shared partners (esp) and non-edgewise shared partners (nsp) need a natural number in order to be well defined \cite{handcock2015package}. Different natural numbers $k$ lead to different network structures. For example kstar(2) models a different network structure than kstar(3). Since we can not consider an infinite, although countable, amount of endogenous variables we need to find an upper bound for the input parameter. Begin by sequentially fitting uni-variate exponential random graph models starting from $k=1$ until we reach $k=N_k$. The value of $N_k$ is determined after achieving three or more consecutive infinite parameter estimates. After finding such $N_k$, we need not consider the consecutive endogenous variables as they will all have coefficient estimates of negative infinity. The results of this procedure will dictate the size of the initial set of endogenous variables denoted by $\mathcal{S}_N$.

\begin{algorithm}
\caption{Bounding the Input Parameter $k$}
\begin{enumerate}
    \item Require an endogenous variable $s_i(y,j) $ with an input parameter $j \in \mathbb{N}$. 
    \item Sequentially fit uni-variate ERGM with with an edge term and endogenous variable $s_{i}(y,j)$. \hspace{1em} Start with $j=2$. 
    \item Store the upper bound number $N_k$ when the parameter estimates for  $s_i(y,N_k + 2)$, $s_i(y,N_k + 1)$ and $s_i(y,N_k )$ all yield negative infinity.
    \item Repeat the above procedure for the following endogenous variables: dyadwise shared partners (dsp), edgewise shared partners (esp), non-edgewise shared partners (nsp) and kstar. Denote the respective upper bounds by $N_{1}$, $N_{2}$, $N_{3}$ and $N_4$.
    \item Obtain a final set of endogenous variables $\mathcal{S}_N$.
\end{enumerate}

\end{algorithm}

\subsection{Stochastic Forward Selection}

After applying the first algorithm for an observed network, one obtains a list of $N$ candidate endogenous variables, $\mathcal{S}_N$, where the size of this set is determined by the upper bounds of dsp, esp, nsp and kstar. Begin by fitting an exponential random graph model with an edge term. The edge term is analogous to the intercept term for a linear model. This null model, also called Bernoulli Random Graph, assumes that the probability of a tie formation between to different nodes follows a Bernoulli distribution. Furthermore, it restrictively assumes that the probability of a tie formation between two nodes is independent of all other ties formed in the network \cite{kolaczyk2014statistical}. In other words, it  does not account for the dependence of observations. The null model will serve as the basis of comparison for candidate ERGMs. We sequentially fit uni-variate ERGMS for each element in  $s_i \in \mathcal{S}_N$ for $i \in \{1,\dots, N \}$. Explicitly the uni-variate ERGMs is formulated below.

\begin{equation}
    P_{\theta}(Y_{i,j}=1) = \psi(\theta_0, \theta_i) exp \{ \theta_0 s_0(y) + \theta_i s_i(y)  \} \hspace{2em} i\in \{ 1,\dots,N \}
\end{equation}

Recall that $Y_{ij}$ is the binary outcome of a whether a tie was formed between two nodes. Here, $y$ is the observed network at a given state. The notorious normalizing constant is given by $\psi$. The edge term and it's associated regression coefficient is given by $s_0(y)$ and $\theta_0$ respectively. The endogenous variable under consideration is $s_i(y)$ along with it's regression coefficient $\theta_i \in \mathbb{R}$. Denote the null ERGM's AIC by $AIC_0$ and each of the $N$ uni-variate ERGMs' AIC by $AIC_i$ for $i \in \{1,\dots,N\}$. The decision of whether to select the endogenous variable, $s_i(y)$, will hinge on the quantity below.

\begin{equation}
    b_i = \frac{AIC_0 - AIC_i}{AIC_0} \hspace{1em} i \in \{ 1, \dots, N \}
\end{equation}

\vspace{1em}

 The main idea of the second algorithm is that an endogenous variable $s_i(y) \in \mathcal{S}_N$ is only discarded if it consistently increases the AIC when compared to the null model. However, estimating the likelihoods of the above models via Monte Carlo Markov Chain (MCMC) is troublesome as poor initial parameter estimates can lead to model divergence and significantly increase computational time \cite{geyer1992constrained}. This motivates the use of Contrastive Divergence (CD) to seed initial parameter estimates via importance sampling to perform Monte Carlo Maximum Likelihood Estimation (MC MLE) developed by Krivitsky \cite{krivitsky2017using}. The increase in computational efficiency comes at a cost, the proposed models' likelihoods and thus their AIC now randomly fluctuate. The fluctuation in the model's likelihood estimate prevents us from using the typical forward selection methods of generalized linear models. There are two types of fluctuations that occur, the first fluctuation mimics the behavior of the error term of a regression model in the sense that is Gaussian. The second fluctuation arises from poor initial starting values and leads to a phenomena known as ERGM degeneracy \cite{li2015degeneracy}. \newline 

Due to this fluctuation, it is possible for a candidate endogenous variable to explain an observed network's topology yet increase the AIC. This increase in AIC is not due to the endogenous variable's predictive power but rather a poor initial parameter guess. To prevent discarding potentially informative endogenous variables which reveal an observed network's structures, we proposed a more flexible forward selection procedure that is based on the 10th percentile of the relative AIC change given by $\hat{b}_i$.

\begin{algorithm}
\caption{Stochastic Forward Selection for Endogenous Variables}
\begin{enumerate}
    \item Require a set $\mathcal{S}_N = \{ s_1(y), \dots, s_N(y) \}$ consisting of $N$ candidate endogenous variables.
    
    \item Fit a null model ERGM with just an edge term. Denote the AIC value by $AIC_0$.
    
    \item For $i \in \{1, \dots, N \}$ : 
        \begin{itemize}
            \item Sequentially fit uni-variate ERGMs with one endogenous variable at a time; $P_{\theta}(Y_{i,j}=1) = \psi(\theta_0, \theta_i) exp \{ \theta_0 s_0(y) + \theta_i s_i(y) \} $.
            \item Record the estimate of the AIC for each uni-variate ERGM by $AIC_i$ and  compute the relative AIC change $b_i$. 
            \item Refit the uni-variate ERGM $M$ times and compute the 10th percentile of the relative AIC change denoted by $\hat{b}_i$.           
        \end{itemize}

        \item  \underline{If} $\hat{b}_i \leq 0$ then remove $s_i(y)$ from the set $\mathcal{S}_N$.  
\end{enumerate}
\end{algorithm}

\subsection{Categorizing the Endogenous Variables based on relative AIC change}

There exists three categories which an endogenous variable $s_i(y) \in \mathcal{S}_N$ can fall under. The first category, denoted by $D_1$, suggests that we should include this variable in our model. Endogenous variables in this category produce a normally distributed relative AIC change that is always positive. This occurs when the proposed endogenous variable always decreases the  AIC when compared to the null ERGM. The second category, denoted by $D_2$, indicates that we should not consider the endogenous variable to model a given network. There are two possibilities for the distribution of the relative AIC change in this scenario. First, the endogenous variable under consideration produces degenerate ERGMs which yields a very negative relative AIC change. Second, the endogenous variable produces a "stable" relative AIC change which is normally distributed yet has a mean relative AIC change considerably less than 0. \newline 
 
The third category, denoted by $D_3$, occurs when inclusion of the endogenous variable sometimes leads to an increase or decrease of the AIC with respect to the null ERGM. For endogenous variables that fall in this category it is unknown whether the increase in relative AIC is due to poor initial parameter seeds or if the variable is not predictive of the observed network's structure. This motivates the use of the tenth percentile for the relative AIC change, $\hat{b}_i$. If $\hat{b}_i$ is positive then this is strong evidence that the endogenous variable is predictive of network structure. On the other hand, if $\hat{b}_i$ is negative then this indicates that the endogenous variable under consideration is not predictive of network structure.

\subsubsection{An illustrative Example for Endogenous Variables in $D_3$}

This example serves to elucidate the need for the relative AIC percentile in the stochastic forward selection algorithm. We applied algorithm 2 to a transcription regulation network for Ecoli \cite{hummel2010steplength} \cite{salgado2001regulondb} \cite{shen2002network}. The relative AIC change between the null ERGM and the univariate ERGM was recorded $M=90$ times. The uni-variate ERGM under consideration contained the endogenous variable degree cross product. The density of the relative AIC change is found in figure 2. Of the 90 times the relative AIC was recorded, 5 of them resulted in a huge increase in AIC when compared to the null model. The 10th percentile of the relative AIC for degree cross product is positive with a median value 0.0265. Had we only calculated the relative AIC change once, it would have been possible to falsely exclude the degree cross product variable. As shown in figure 6, this endogenous variable was predictive of the observed network's structure.

\begin{figure}[ht]
    \centering
\includegraphics[width=13cm]{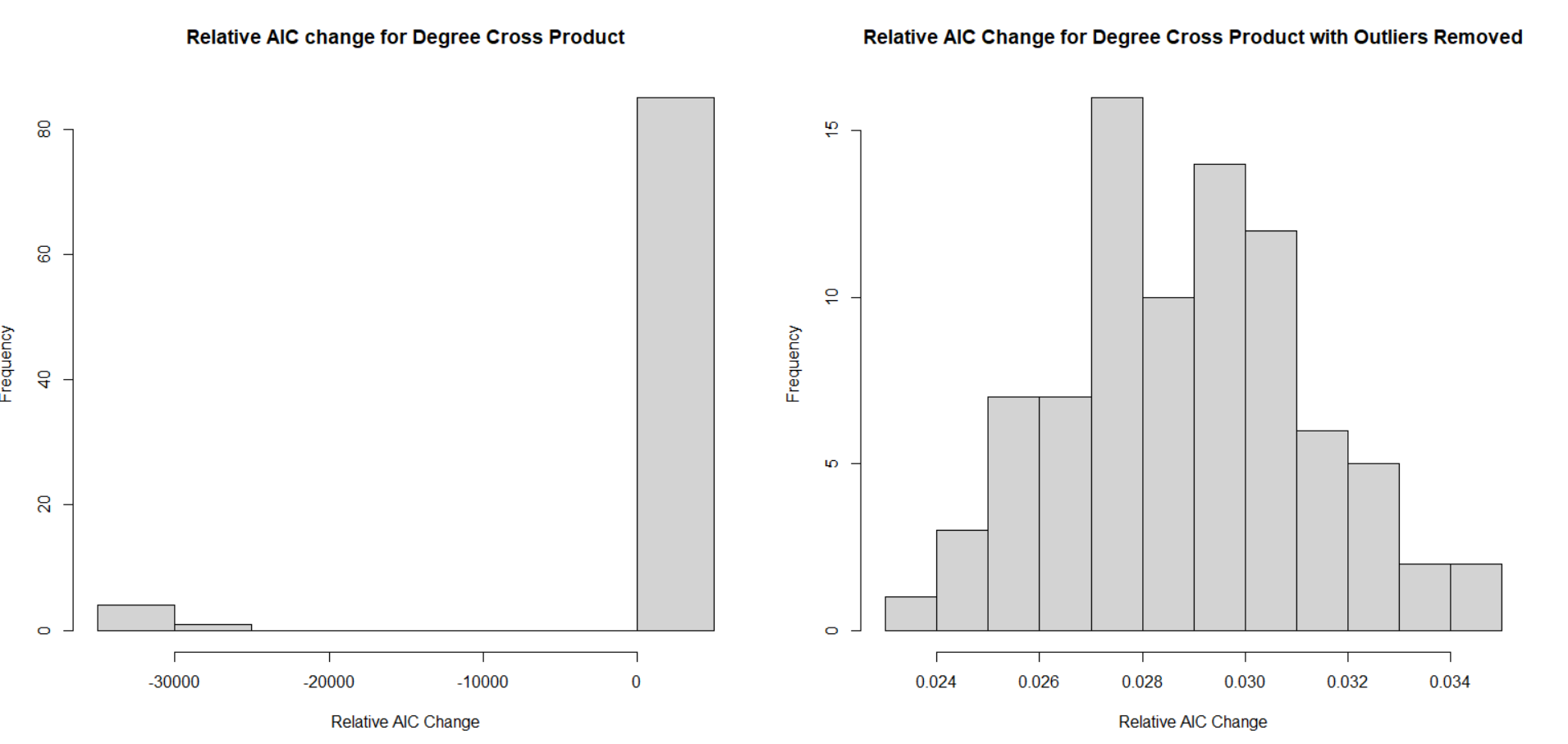}
    \caption{ Frequency of the relative AIC change for the uni-variate ERGM with degree cross product as the main predictor. The histogram on the left is an example of relative AIC fluctuation due to poor initial starting points. The histogram on the right exhibits relative AIC fluctuation that mimics random noise. }
\end{figure}

\subsection{ Degeneracy Screening }



Exponential Random Graph models with multiple endogenous variable have a tendency to exhibit multi col-linearity which increases the probability of simulating degenerate networks \cite{li2015degeneracy}. This col-linearity is detrimental to the overall performance of the model. In an effort to reduce this potential correlation, we will limit proposed ERGMs to have at most three endogenous variables including the intercept or edge variable. Furthermore, this restriction allows the researcher to comfortably add exogenous variables without the risk of running into model over-fit.  Denote the set of all possible models by $\mathcal{M}$. \newline 

Model selection will be based on network motif counts denoted by $H$. We present a flexible guideline to perform model selection via homomorphism densities. For an observed network, there exists multiple sampling methods to generate homomorphism densities \cite{bhattacharyya2015subsampling} \cite{naulet2021bootstrap}. Denote the set of $n_{\mathcal{H}}$ networks motifs by $\mathcal{H} = \{ H_1, H_2, \dots, H_{n_{\mathcal{H}}}  \}$. Denote the observed network by $O$ and the proposed ERGM by $M \in \mathcal{M}$. 


\begin{algorithm}
\caption{Degeneracy Screening Algorithm}
\begin{enumerate}
    \item Require a non empty network motif set  $\mathcal{H}$. 
    \item For an observed network $O$, compute the homomorphism densities $t(H_i,O)$ for each  network motif $H_i \in \mathcal{H}$ for  $i \in \{1, \dots , n_{\mathcal{H}} \}$. 
    \item For a proposed ERGM  $M\in \mathcal{M}$, compute the homomorphism densities $t(H_i,M)$ for each  network motif $H_i \in \mathcal{H}$ for  $i \in \{1, \dots , n_{\mathcal{H}} \}$.
    \item Define a summary statistic $S_i(t)$, to be applied to the homomorphism densities $t$ for $i \in \{1, \dots, n_{\mathcal{H}} \}$.
    \item Compute the difference $S_O(t(H_i,O) - S_M(t(H_i,M)$.

    \item If $|S_O(t(H_i,O) - S_M(t(H_i,M)| \geq c_{S_i} $ then discard $M$ from the set of possible models.
\end{enumerate}
\end{algorithm}

\subsubsection{Network Motifs and Summary Statistics Used}

In this paper, the summary statistics for the proposed model $M$ and the observed network $O$ is the first moment. The network motifs used were edges, 2-stars and triangles. Model selection was based on how close the mean number of edges, 2-stars and triangles were to the observed network count of edges, 2-stars and triangles. The models selected for the result visualizations in the appendix had the closest number of edges, 2-stars and triangles compared to the observed network. The use of homomorphism densities in algorithm 3 allows the researcher to select ERGMs in a flexible manner. By using the appropriate summary statistic, the researcher is able to select ERGMs that capture particular community structures found in the observed network. For example, consider an observed network with 5 isolated nodes as in the Florentine business network. Algorithm 3 allows the researcher to select ERGMs that mimic this community of isolated nodes. Let the network motif be an isolated node and choose the summary statistic to be the minimum count of isolated nodes. By setting $c_{S_i}$ to be 1, the researcher can screen out candidate ERGMs that do not possess this isolated community structure.

\subsection{Ensuring a Stable ERGM proposal}

The Hamiltonian operator $H(Y)$ plays a central role in limiting the degeneracy phenomena exhibited by ERGMs. For a tie, $e$, one can decompose the operator into $H(Y) = A_e(Y) + B_e(Y)$ where $Y$ is our network, $A_e(Y)$ consists network structures dependent on the tie $e$ and $B_e(Y)$ consists of network structures independent of the tie $e$. During an ERGM draw, one starts with an initial state $Y_0$ for the network and updates each potential tie in an in an iterative manner. One can characterize the future state $Y_{t+1}$ by either $Y_{e^+}$ or $Y_{e^-}$. Here $Y_{e^+}$ is identical to the network $Y_{t}$ except that it possess an extra tie $e$. Similarly, $Y_{e^-}$ is identical to $Y_{t}$. The transition probability from $Y_t \longrightarrow  Y_{t+1} $ is completely determined by the derivative of the Hamiltonian $H(Y)$ \cite{bhamidi2008mixing}. Explicitly we have the probability of obtaining state $Y_{e^+}$ is given by $\frac{exp\{ \partial H(Y) \}}{ 1 + exp\{ \partial H(Y) \}}$. Thus, the probability of generating degenerate networks from a proposed ERGM is intimately related to the smoothness properties of the Hamiltonian operator $H(Y)$ and the behavior of it's derivatives. In essence, the procedures of this paper choose the Hamiltonian in an "informed" way thus significantly reducing the probability of model degeneracy. 

\subsubsection{Role of Null ERGM}

The null ERGM, also known as the Bernoulli random graph, is guaranteed to produce a non-degenerate ERGM \cite{butts2011bernoulli}. ERGM degeneracy can be characterized by a very large AIC value \cite{li2015degeneracy}. Since algorithm two uses the null ERGM as the basis of comparison, by choosing endogenous variables that consistently decrease AIC compared to the null ERGM, we guarantee non-degenerate uni-variate model proposals. However, when we consider bi-variate ERGMs, we run the risk of inducing model degeneracy caused by col-linearity. This motivates the two fold use of algorithm 3 as both a model selection and degeneracy screening procedure. 


\section{Numerical Studies}

To test the potential and limitations of this method, we applied the algorithms of this paper to eleven real life networks \cite{handcock2015package} \cite{caimo2012bergm} \cite{shen2002network} \cite{lazega2001collegial} \cite{kapferer1972strategy} \cite{padgett2011marriage} \cite{resnick1997protecting}. The nodes vary from 16 to 418. All 11 networks are un-directed and posses a binary outcome value. The 11 networks were split into 3 main categories. The first category represents small networks with nodes at most 20 with total number of edges not exceeding 30. The Florentine marriage, Florentine business, Moody contact and Molecule network belong to this category. The second category of networks consisted of more complicated and slightly larger networks. The total number of nodes ranged from 20 to 80 and the total number of edges ranged from 50 to 200. The Lazega lawyer network, Kapferer 1 and 2 tailor shop network , Zach karate networks belonged to this category. Finally, the last category of networks possessed the largest and most complicated network structures. The number of nodes ranged from 80 to 418 with the total number of edges ranging from 200 to 556.

\subsection{Bounding the Input Parameter:}

The results of the first algorithm provide us with an initial model space of ERGMs for each of the 11 networks. The size of this set is determined by the results of algorithm 1. The model space consists of ERGMs with two distinct endogenous variables and the edge term as well as one endogenous variable and the edge term. Explicitly written we have ERGMs of the following form.

\begin{equation}
    P_{\theta}(Y_{i,j}=1) = \psi(\theta_0, \theta_i) exp \{ \theta_0 s_0(y) + \theta_i s_i(y)  \} \hspace{2em} i\in \{ 1,\dots,N \}
\end{equation}

\begin{equation}
    P_{\theta}(Y_{i,j}=1) = \psi(\theta_0, \theta_i, \theta_j) exp \{ \theta_0 s_0(y) + \theta_i s_i(y) + s_j(y) \} \hspace{1em} i, j\in \{ 1,\dots,N \} \hspace{1em} i \neq j
\end{equation}

The edge term is represented by $s_0$, the distinct endogenous variable are represented by $s_i$ and $s_j$. The size of the model space for these 11 different networks varies from 153 to 9316 different ERGMs. The number of possible models is directly related to the number of endogenous variables as there exists a one-to-one function mapping the two sets to one another.

\begin{table}[ht]
\resizebox{14cm}{!}
{
 \begin{tabular}{||c c c  ||} 
 \hline
   Network: & $\#$  of Endogenous Variables: & $\#$ of Possible Models  \\ [0.5ex] 
 \hline\hline 
Lazega & 44  &  990  \\ 
 \hline
  Kapferer & 62  & 1953   \\ 
 \hline
  Kapferer2 & 63  &  2016  \\
 \hline
   Zach &  43  &  946   \\
 \hline
    Wind Surfers & 136  & 9316    \\
 \hline
   Molecule &  17 & 153   \\
 \hline
   Faux Mesa High school &  32 & 528    \\
 \hline
 Ecoli &  101 &  5151  \\
 \hline
  Moody Contact Sim & 17 & 153   \\
 \hline
 Florentine Marriage& 18  & 171    \\
 \hline
 Florentine Business & 17 &  153 \\  
 \hline
\end{tabular}
}
    \caption{ This table contains the initial order of the model space for each network via the number of endogenous variables and number of possible models.}
\end{table}

\begin{table}
\resizebox{14cm}{!}{
 \begin{tabular}{||c c c c c ||} 
 \hline
   Network: & Upper Bound of kstar & Upper Bound of esp & Upper Bound of nsp & Upper Bound of dsp  \\ [0.5ex] 
 \hline\hline 
 Lazega & 14  &  6 & 7 & 7 \\ 
 \hline
  Kapferer & 23  &  11 & 7 & 11 \\ 
 \hline
  Kapferer2 & 24  & 12  & 7 & 12\\
 \hline
   Zach & 16  & 9  & 5 & 9 \\
 \hline
    Wind Surfers &  57 & 35  & 15 & 35\\
 \hline
   Molecule &  4 &  1 & 1 & 1\\
 \hline
    Faux Mesa High School &  12  & 4 & 3 & 4 \\
 \hline
 Ecoli & 71 & 6  & 9 & 9\\
 \hline
  Moody Contact Sim & 5  & 0   &  1& 1 \\
 \hline
 Florentine Marriage &  5  & 1   & 1 & 1 \\
 \hline
 Florentine Business&  4 & 1 & 1 & 1\\  
 \hline
\end{tabular}
}
\caption{ This table contains the upper bounds for kstar, edgewise shared partners (esp), non-edgewise shared partners (nsp) and dyadwise shared partners (dsp).}
\end{table}

\newpage 

\subsection{Stochastic Forward Selection:}



In an effort to reduce the order of the initial model space, we applied the stochastic forward selection algorithm to each of our 11 networks. The initial model space was consistently reduced by an order of magnitude across all  networks.

\begin{table}[ht]
\resizebox{14cm}{!}{
 \begin{tabular}{||c c c c ||} 
 \hline
   Network: & $\#$  of Endogenous Variables: & $\#$ of Possible Models & Endogenous Variables in $D_3$  \\ [0.5ex] 
 \hline\hline 
  Lazega & 10  & 55  & 1 \\ 
 \hline
   Kapferer & 12  & 78  & 1  \\ 
 \hline
   Kapferer2 & 13  & 91  & 3  \\
 \hline
    Zach & 6   &  21 & 1  \\
 \hline
     Wind Surfers & 25  &  325 & 7  \\
 \hline
    Molecule & 9  & 45  & 0  \\
 \hline
     Faux Mesa High School & 7  & 28 & 1  \\
 \hline
  Ecoli & 14   & 105   &  3\\
 \hline
   Moody Contact Sim & 7  & 28   & 7  \\
 \hline
  Florentine Marriage &  2  & 3  & 1  \\
 \hline
  Florentine Business & 3 & 6 & 2  \\  
 \hline
\end{tabular}
} 
    \caption{This table contains the  order of the model space after the endogenous variable feature selection algorithm was applied for each network. }
\end{table}

Based on the the results of the initial screening, one can create a metric of network complexity that is based on the number of possible models proposed by this algorithm. For example, despite having the same number of nodes at 16 and similar number of edges that range between 15 and 20, the Moody contact network possess 28 candidate models while the Florentine business and marriage networks possess just 6 and 3 candidate models respectively. The Moody contact network is originally directed and both the Florentine networks are un-directed. Despite simplifying the Moody contact network by replacing the directional ties, it still retains some complexity as illustrated by the size of the model space. Similarly, the number of endogenous variables in $D_3$ can also be used as a metric of network complexity. Recall, $D_3$ is the set of endogenous variables which decrease the model's AIC compared to the null but sometimes increases it due to poor initial parameter seeds. Again, looking at these three networks, we see that the Moody contact network posses 7 endogenous variables in $D_3$ while the Florentine marriage and business network only posses 1 and 2 respectively. The Kapferer and Kapferer2 network are friendship networks of a tailor shop in Zambia observed at two distinct time points. At the second time point, Kapferer2 is more complex as there are 4 additional workers and 32 additional ties. The size of the model space and set $D_3$ illustrate this added complexity. Indeed, the model space increased from 78 to 91 and Kapferer2 possesses 2 extra endogenous variables in $D_3$.

\vspace{1em}

The reduction of the model space by an order of magnitude allows us to consistently propose ERGMs for small un-directed networks such as the Florentine marriage network. Below is a visualization of the ERGM proposed via this method for the Florentine marriage network with endogenous variables only. The endogenous variables used were the kstar(2) and kstar(3) terms. Below is a visualization of three networks simulated from the ERGM.

\begin{equation}
        P_{\boldsymbol{\theta}}(Y_{i,j}=1) = \psi({\boldsymbol{\theta}}) exp \{ \theta_0 \times  {edges} + \theta_1 \times kstar(2) + \theta_2 \times kstar(3) \}
\end{equation}

\begin{figure}[ht]
    \centering
\includegraphics[width=12cm]{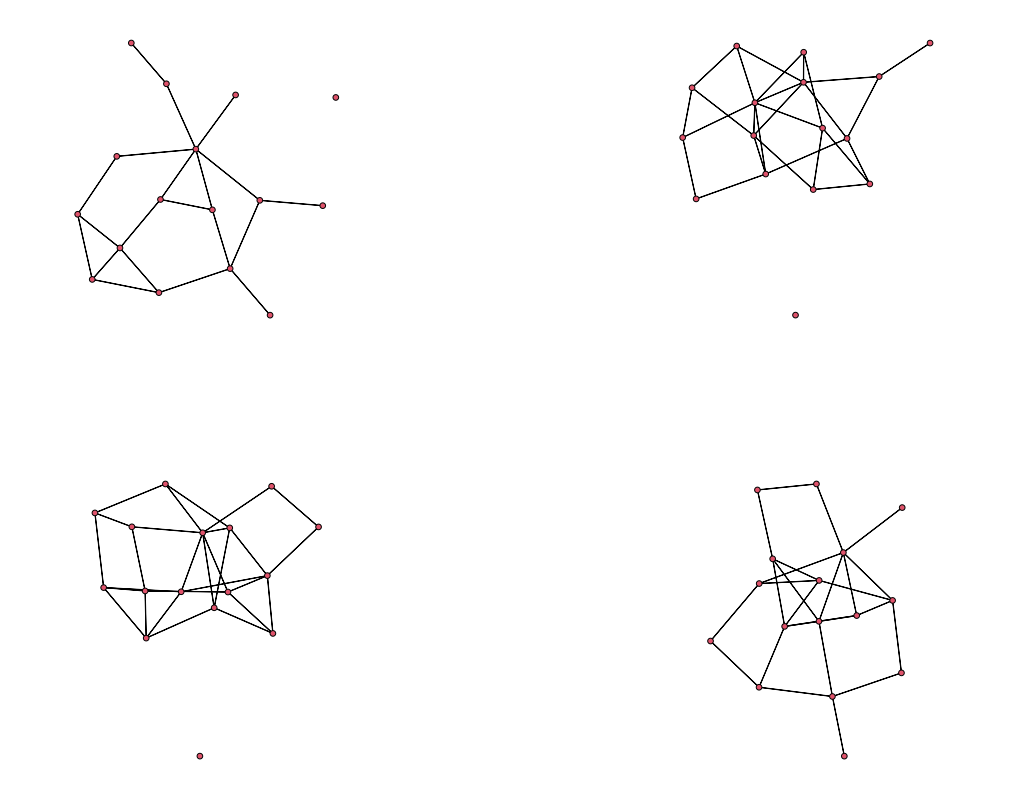}

    \caption{ Upper left network denotes the observed Florentine marriage network. The remaining three networks are draws from the ERGM in equation 13 produced via algorithm 1 and 2 }
    \label{molecule_kstar(3)_and_kstar(2)}
\end{figure}

\newpage

\subsection{Degeneracy Screening and Model Selection Results}

Model selection was based on the counts of edges, 2-stars and triangles found in the observed networks. Algorithm 3 was used for degeneracy screening and to pinpoint the best performing ERGMs. Degenerate ERGMs possessed an average count of edges that deviated greatly from the edge homomorphism (tie counts) densities of the observed network. The counts of edges, 2-stars and triangles are reported below for each observed network. The average count of edges, 2-stars and triangles for the ERGMs proposed by the methods of this paper are reported in the table below. 

\begin{table}[ht]
\resizebox{14cm}{!}{
 \begin{tabular}{||c c c c c ||} 
 \hline
   Network: & Number of Nodes: & Number of Edges: & Number of 2-stars &  Number of Triangles :  \\ [0.5ex] 
 \hline\hline 
   Lazega & 36  & 115 & 1852 & 240 \\ 
 \hline
    Kapferer & 39  & 158 & 3132 & 402 \\ 
 \hline
    Kapferer2 & 43  & 190  & 4074 & 504 \\
 \hline
     Zach & 34  & 78  & 1056 & 90\\
 \hline
      Wind Surfers* &  95 & 556  &  21966 & 2890 \\
 \hline
     Molecule & 20  & 28  & 120 & 12 \\
 \hline
      Faux Mesa High School& 205  & 203  &  1318 & 124\\
 \hline
   Ecoli* & 418  & 519  & 10580 & 84 \\
 \hline
    Moody Contact Sim* & 16  & 18 & 80  & 0 \\
 \hline
 Florentine Marriage & 16  & 20 & 94 & 6\\
 \hline
 Florentine Business & 16 & 15 & 72  & 10 \\  
 \hline
\end{tabular}
}
    \caption{ Network motif counts for the 11 different networks used in our investigation. The * indicates that the network posses a directed counterpart.}
\end{table}

\begin{table}[ht]

\resizebox{14cm}{!}{
 \begin{tabular}{||c c c c c ||} 
 \hline
   Network: & $\#$ of Models & Average $\#$ of Edges: & Average $\#$ of 2-stars & Average $\#$ of Triangles :  \\ [0.5ex] 
 \hline\hline 
   Lazega & 85   & 224.03 & 1924 & 174.2 \\ 
 \hline
    Kapferer &  128 & 288.8 & 2387 & 207.14 \\ 
 \hline
    Kapferer2 & 150 &  293.2  & 2792 & 222 \\
 \hline
     Zach & 16  & 117.6  &  1061 & 81.3 \\
 \hline
      Wind Surfers &  324 & 1043  & 12741  & 770 \\
 \hline
     Molecule & 69  &  29.68 & 123.4 & 3.98 \\
 \hline
      Faux Mesa High School & 39  & 344.4  &  754.3 & 96.90 \\
 \hline
   Ecoli & 104  & 869   & 2057 & 9.54 \\
 \hline
    Moody Contact Sim &  37 & 56.24 &  79.46 & 3.628 \\
 \hline
   Florentine Marriage &  1 & 24.97 & 147.5  & 12.59 \\
 \hline
   Florentine Business & 2 & 22.80  & 59.13  & 23.93 \\  
 \hline
\end{tabular}
}
    \caption{ Counts of the network motifs for the ERGMs proposed by our methods for the 11 different networks. Models are comprised of ERGMs with endogenous variables only. }
    \label{fig3}
\end{table}


For this reduced model space,  the average number of edges, 2-stars and triangles tend towards the observed count. The number of edges were overestimated for all networks. This bias is inherent to exponential family models \cite{efron1978geometry}. On the other hand, the average number of 2-stars and triangles were sometimes above the observed count and below the observed count for different networks. The exclusion of exogenous variables from these candidate ERGMs explains the discrepancy between the observed network motif counts and the counts generated by the ERGMs. Recall, in an ERGM nodal attributes and their influence on network structure are modeled via exogenous variables. A remedy to this phenomenon that does not require incorporating the appropriate exogenous variables is the use curved ERGMs or allowing the candidate ERGM to possess more than 2 endogenous variables\cite{hunter2007curved}. 

\begin{table}[ht]
\resizebox{10cm}{!}{
 \begin{tabular}{||c c c ||} 
 \hline
   Network: & $\#$ of Models& $\#$ of Degenerate Models  \\ [0.5ex] 
 \hline\hline 
   Lazega &  50   & 5 \\ 
 \hline
    Kapferer &  74  & 4 \\ 
 \hline
    Kapferer2 &  85     & 6\\
 \hline
     Zach & 16     & 5\\
 \hline
      Wind Surfers & 324     & 1 \\
 \hline
     Molecule &  42   & 3 \\
 \hline
      Faux Mesa High School & 25    & 3\\
 \hline
   Ecoli &  104  & 1 \\
 \hline
    Moody Contact Sim & 23     & 5 \\
 \hline
   Florentine Marriage &3      &0  \\
 \hline
   Florentine Business & 5    & 1 \\  
 \hline
\end{tabular}
}

    \caption{ Number of models after algorithm 3 is applied for degeneracy screening is presented in column one. Number of degenerate models removed is presented in the second column.  }
    \label{modelselection}
\end{table}


\subsection{Nodal Attributes}


Exogenous variables were excluded from the feature selection method as each network possesses different characteristics. We left it to the researcher to decide which exogenous variables to include in an observed network as their inclusion does not induce degeneracy \cite{li2015degeneracy}. As such, none of the papers reviewed by the authors complained about exogenous variable selection \cite{fattore2010network} \cite{okamoto2015scientific} \cite{heaney2014multiplex} \cite{mcglashan2019collaboration}. The inclusion of exogenous variable can be very influential to the structure of the simulated networks as figure 4 and 5 suggest. For the best results, one needs to use both the endogenous variables selected via the methods of this paper and informative exogenous variables based on domain specific knowledge. Note, if only exogenous variables  are used then there is a high probability of losing some of the structures present in the observed network structure as illustrated in figure 18 of the appendix where only exogenous variables were used to model the Florentine marriage network.

\begin{figure}[ht]
    \centering
\includegraphics[width=13cm]{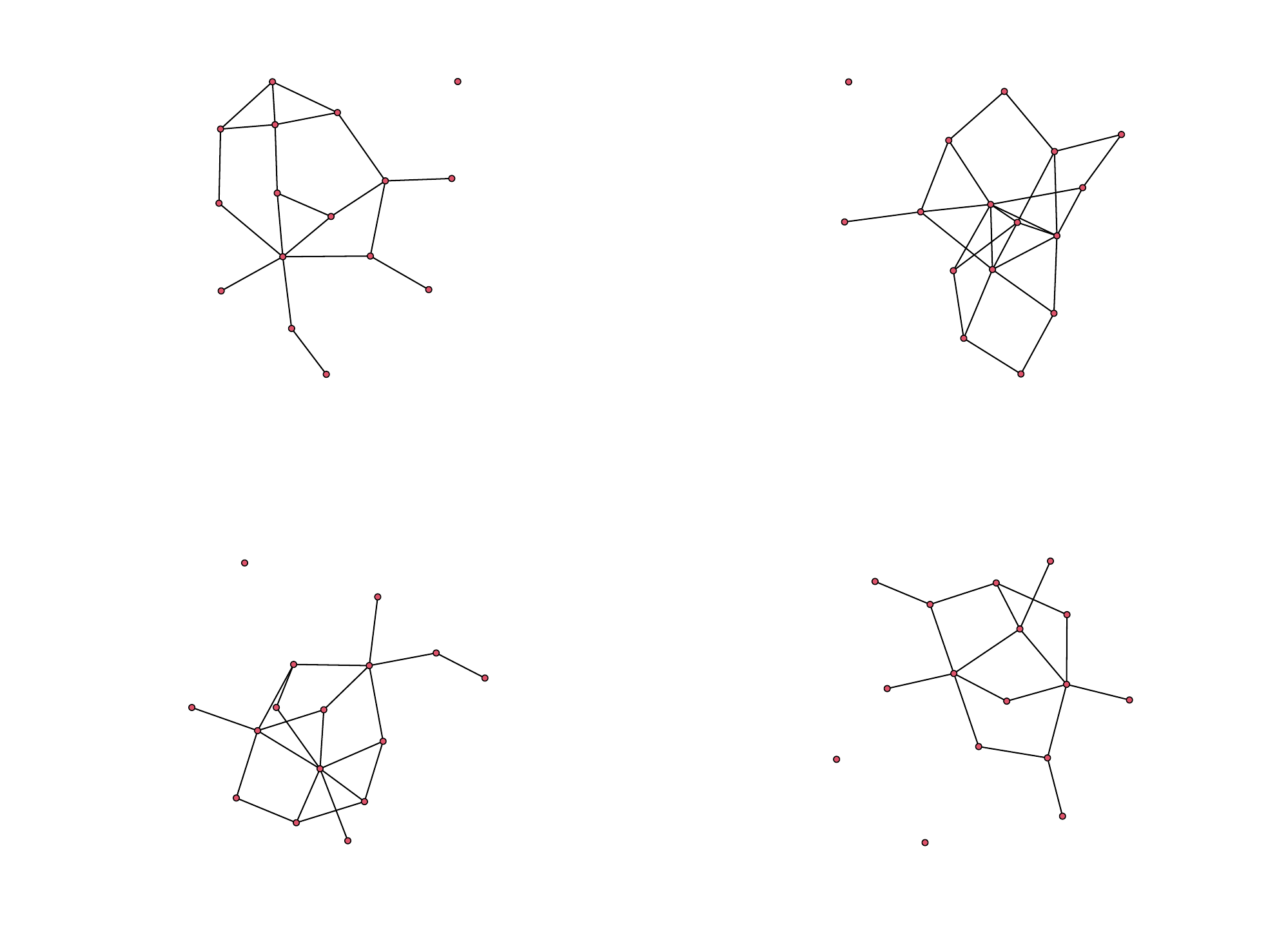}

\caption{ Upper left network denotes the observed Florentine marriage Network. The remaining three networks are draws from the ERGM in equation $            P_{\boldsymbol{\theta}}(Y_{i,j}=1 | X) = \psi({\boldsymbol{\theta}}) exp \{ \theta_0 \times edges + \theta_1 \times kstar(2) + \theta_2 \times kstar(3) + \theta_3 g_1(x) + \theta_4 g_2(x) + \theta_5 g_3(x) \}$. $g_1(x)$ denotes the exogenous variable of familial wealth, $g_2(x)$ denotes the number of seats on the civic council and $g_3(x)$ denotes the total number of business and marriage ties.}
    \label{molecule_kstar(3)_and_kstar(2)_exo}
\end{figure}

\newpage 

\begin{figure}[ht]
    \centering

\includegraphics[width=14cm]{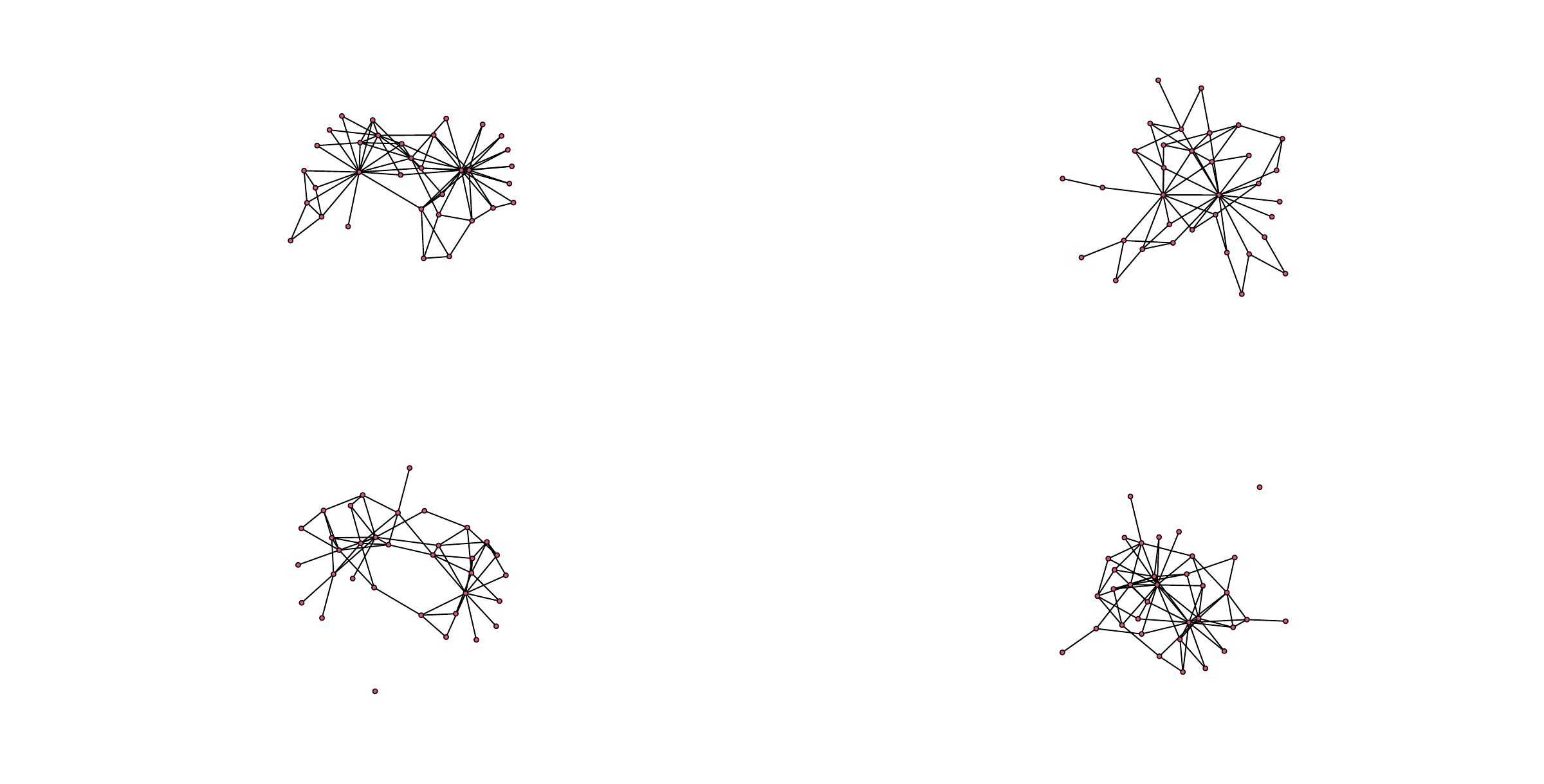}

    \caption{ Upper left network denotes the observed zach karate network. The remaining three networks are draws from the ERGM model with equation $ P_{\boldsymbol{\theta}}(Y_{i,j}=1 | X) = \psi({\boldsymbol{\theta}}) exp \{ \theta_0 edges + \theta_1 esp(3) + \theta_2 dsp(3) + \theta_3 CLUB + \theta_4 FACTION + \theta_5 ROLE \}$. }
\end{figure}

\section{Discussion}
 
The results of the numerical studies section show that the methods of this paper can reliably produce well fitting ERGMs when combined with exogenous variables. The methods of this paper can be used for other types of ERGMs. For example, one can perform variable selection for directed ERGMs by adding endogenous variables that accommodate for directional ties. However, the addition of these directed endogenous variables leads to an inflated model space. For example, the Ecoli network posses 104 candidate ERGMs despite only using un-directed endogenous variables. By performing this extension for other types of ERGMs, one quickly runs into a model selection problem. It is left as an avenue for future research to explore more nuanced model selection procedures for ERGMs to accommodate for this inflated model space. By building off this variable selection procedure, we hope to develop a model selection procedure that allows us to reliably propose ERGMs for larger networks, directed networks and multi-level networks.  \newline

\begin{figure}[ht]
    \centering

\includegraphics[width=13cm]{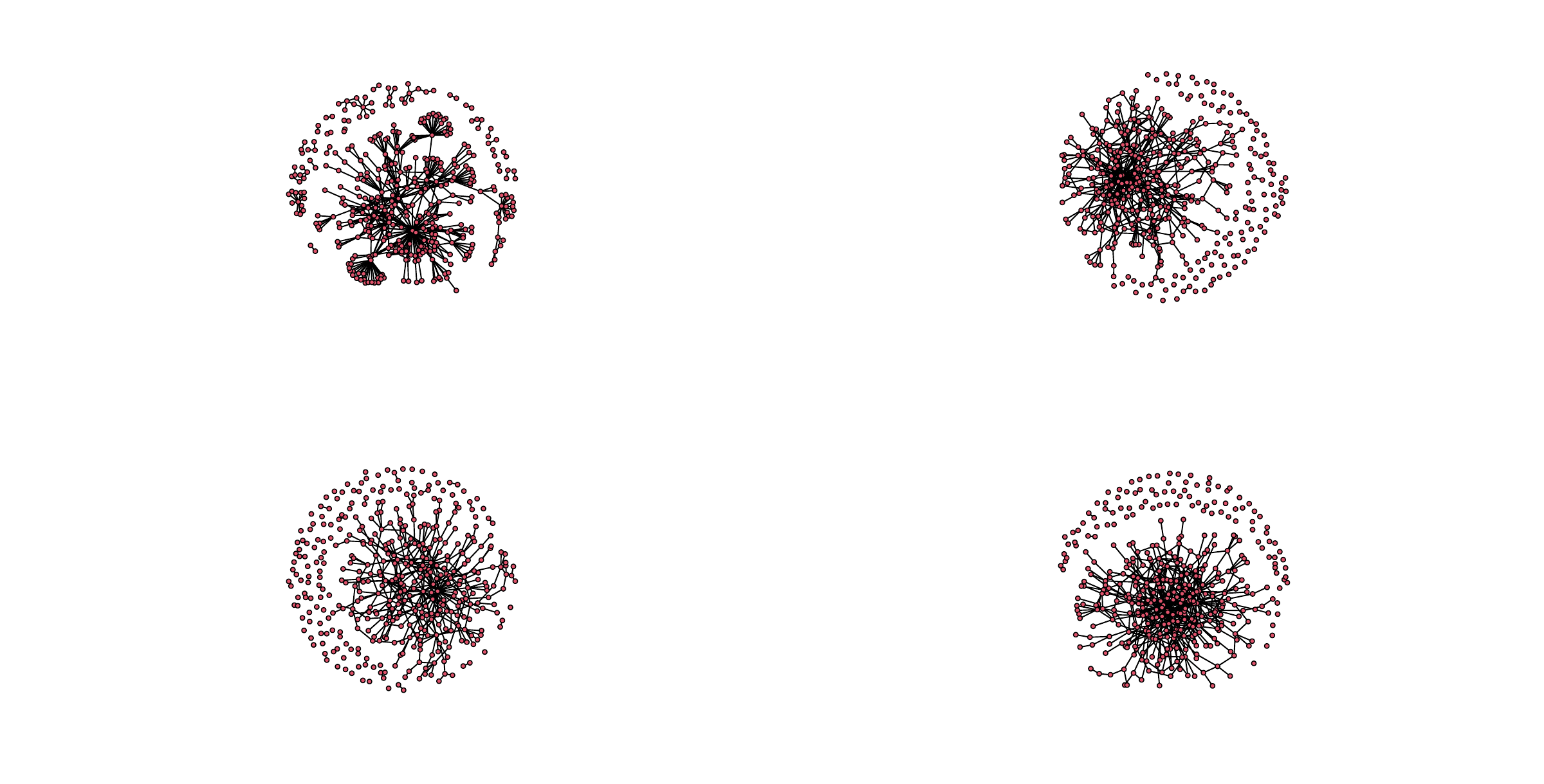}

    \caption{ Upper left network denotes the observed Ecoli network. The remaining three networks are draws from the  ERGM model with equation $ P_{\boldsymbol{\theta}}(Y_{i,j}=1|X) = \psi({\boldsymbol{\theta}}) exp \{ \theta_0 \times edges + \theta_1 dsp(2) + \theta_2 degcrossprod + \theta_3 SELF \}$  }
\end{figure}


Directed networks possess 3 different types of 2 node network motifs and 13 different types of 3 node network motifs. This results in a total of 16 different network motif counts. It becomes apparent that an ERGM scoring method is required to help choose an optimal ERGM for an observed network given the inflated model space. By using principles of the degeneracy screening algorithm, one can construct a model scoring system that complements the selection procedure and highlights different performance aspects of candidate ERGMs. Furthermore, 2 and 3 node network motifs might not be able to capture all the intricacies of the observed network's topology. It is left as an avenue of future research incorporate new metrics that quantify more complicated aspects of an observed network's structure into the model selection and scoring procedures.

\begin{figure}[ht]
    \centering

\includegraphics[width=8cm]{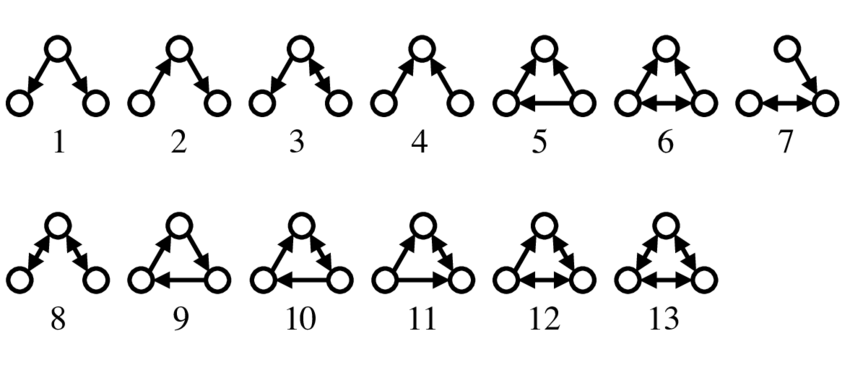}

    \caption{ All 13 possible network motifs with 3 nodes and directed ties. Source:  \cite{bentley2017connectomics}. }
\end{figure}

\bibliographystyle{unsrt}  
\bibliography{references}

\newpage

\appendix
\section{Visualizations}

\begin{figure}[ht]
    \centering

\includegraphics[width=12cm]{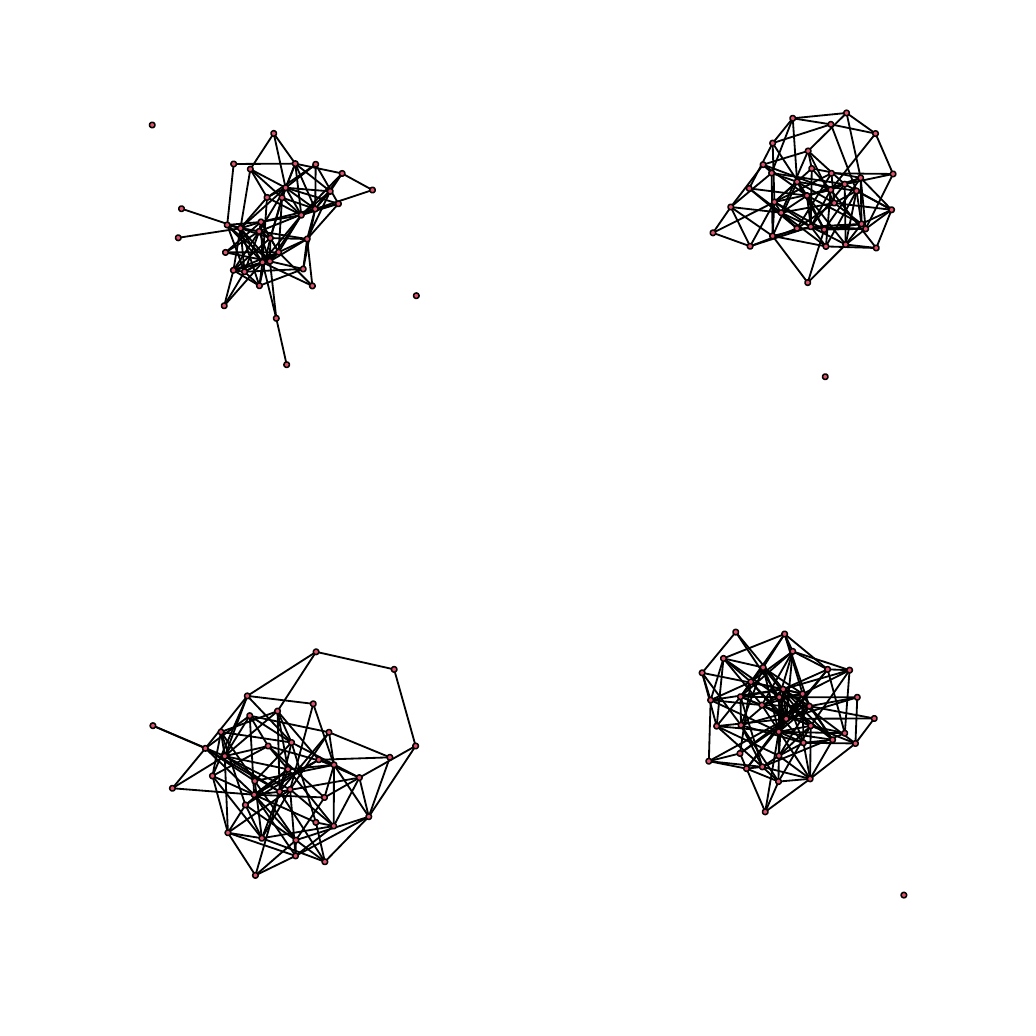}

    \caption{ Upper left network denotes the observed Lazega lawyer network. The remaining three networks are draws from the ERGM model with equation $    P_{\boldsymbol{\theta}}(Y_{i,j}=1) = \psi({\boldsymbol{\theta}}) exp \{ \theta_0 \times edges + \theta_1 nsp(5) + \theta_2 isolates \}$. }
\end{figure}

\newpage



\begin{figure}[ht]
    \centering

\includegraphics[width=15cm]{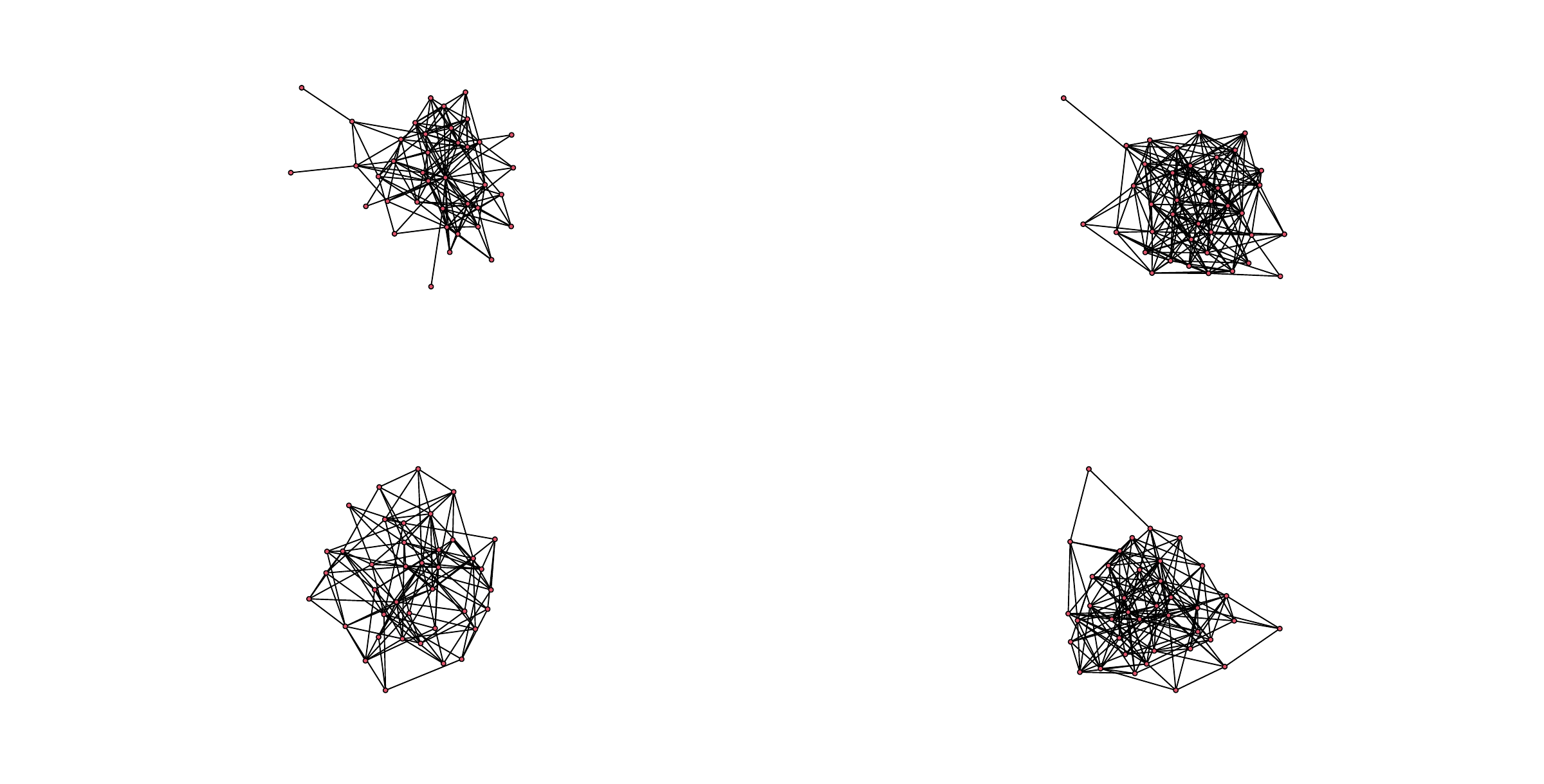}

    \caption{ Upper left network denotes the observed Kapferer network of tailor shop workers. The remaining three networks are draws from the ERGM model with equation $    P_{\boldsymbol{\theta}}(Y_{i,j}=1) = \psi({\boldsymbol{\theta}}) exp \{ \theta_0 \times edges + \theta_1 nsp(5) + \theta_2 dsp(2) \}$.  }
\end{figure}

\begin{figure}[ht]
    \centering

\includegraphics[width=15cm]{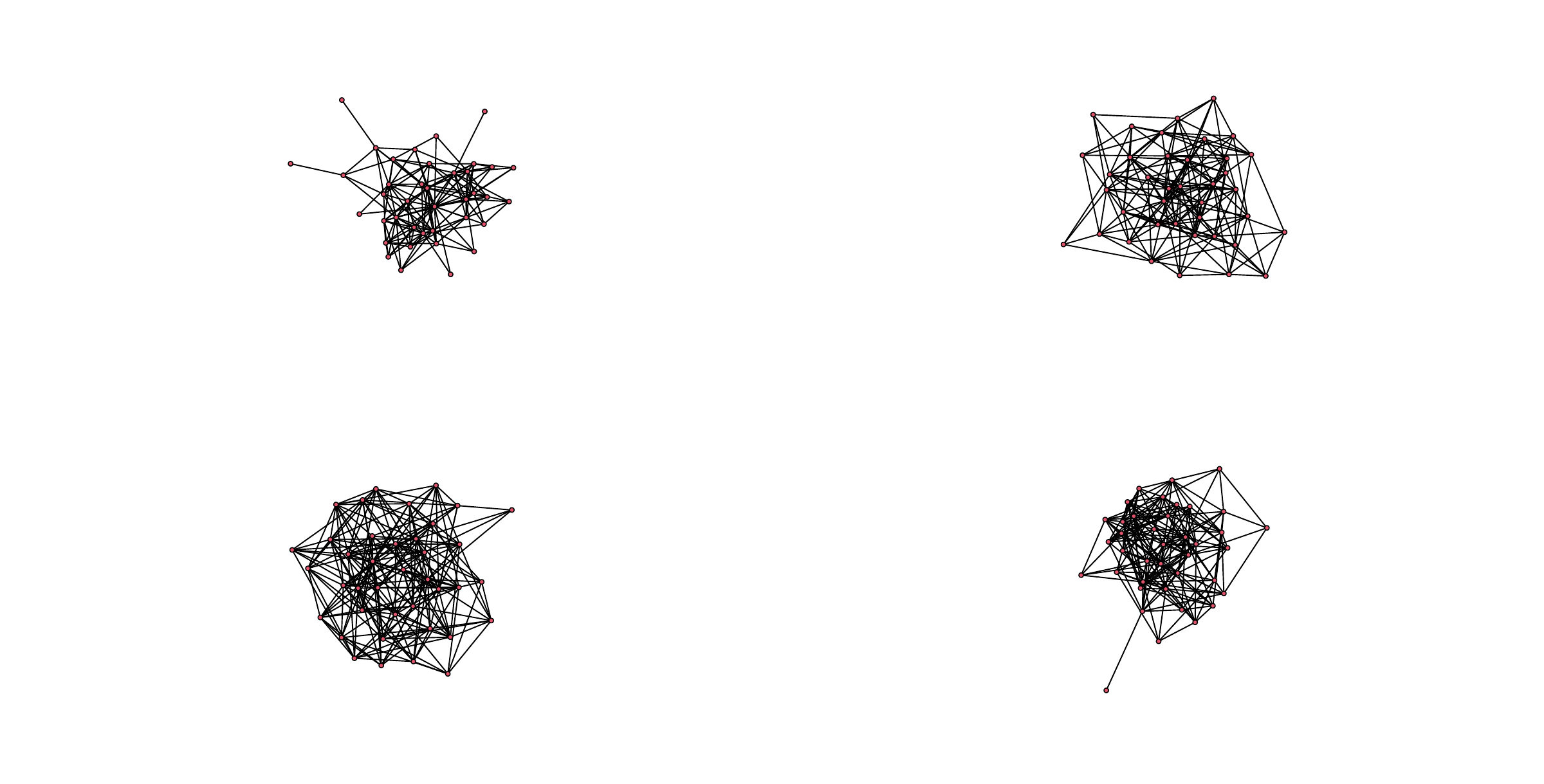}

    \caption{ Upper left network denotes the observed Kapferer network of tailor shop workers. The remaining three networks are draws from the ERGM model with equation $    P_{\boldsymbol{\theta}}(Y_{i,j}=1) = \psi({\boldsymbol{\theta}}) exp \{ \theta_0 \times edges + \theta_1 nsp(5) + \theta_2 dsp(2) \}$.  }
\end{figure}

\begin{figure}[ht]
    \centering

\includegraphics[width=15cm]{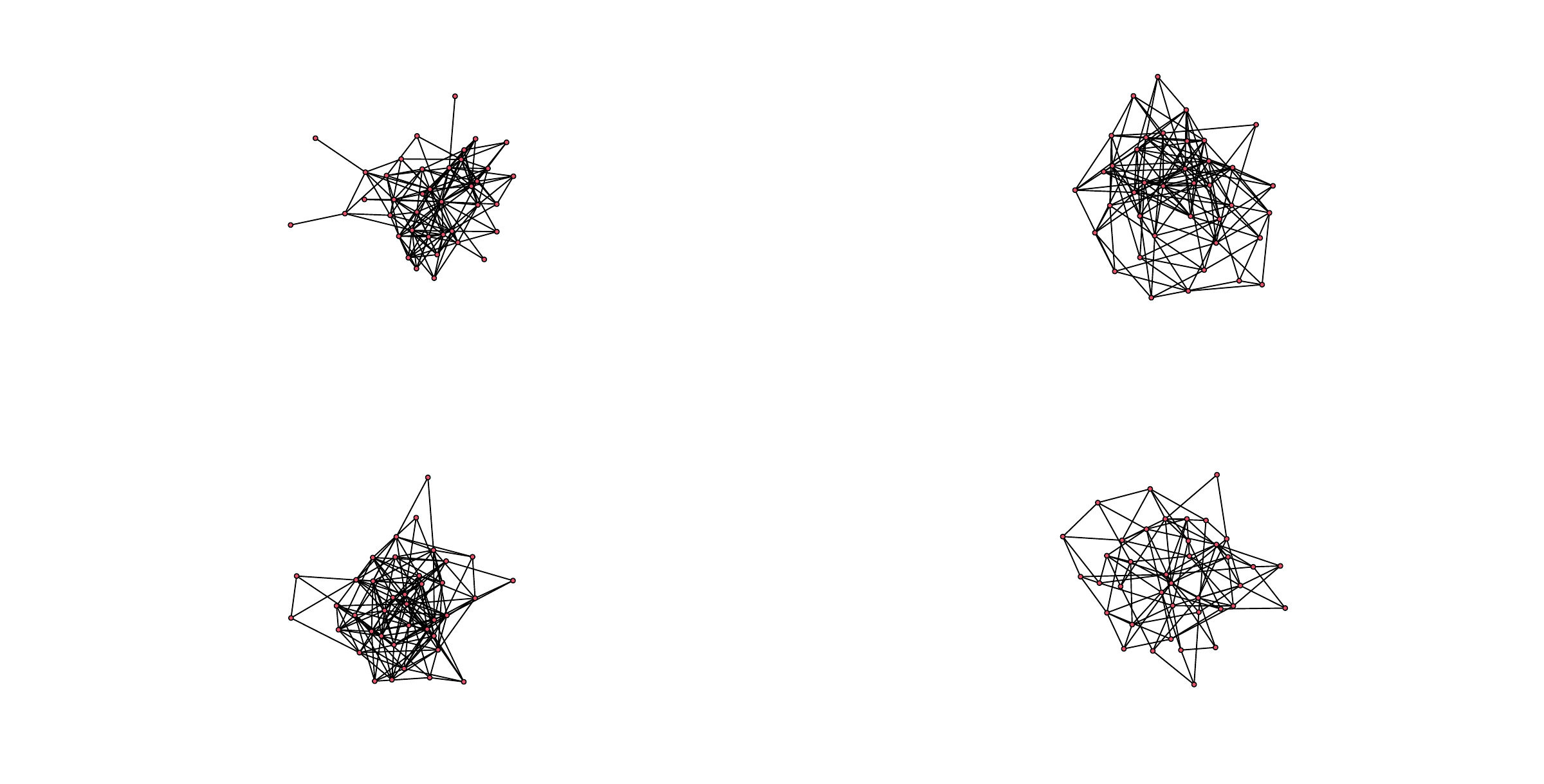}

    \caption{ Upper left network denotes the observed Kapferer network of tailor shop workers. The remaining three networks are draws from the ERGM model with equation $    P_{\boldsymbol{\theta}}(Y_{i,j}=1) = \psi({\boldsymbol{\theta}}) exp \{ \theta_0 \times edges + \theta_1 nsp(5) + \theta_2 dsp(2) \}$.  }
\end{figure}


\begin{figure}[ht]
    \centering

\includegraphics[width=15cm]{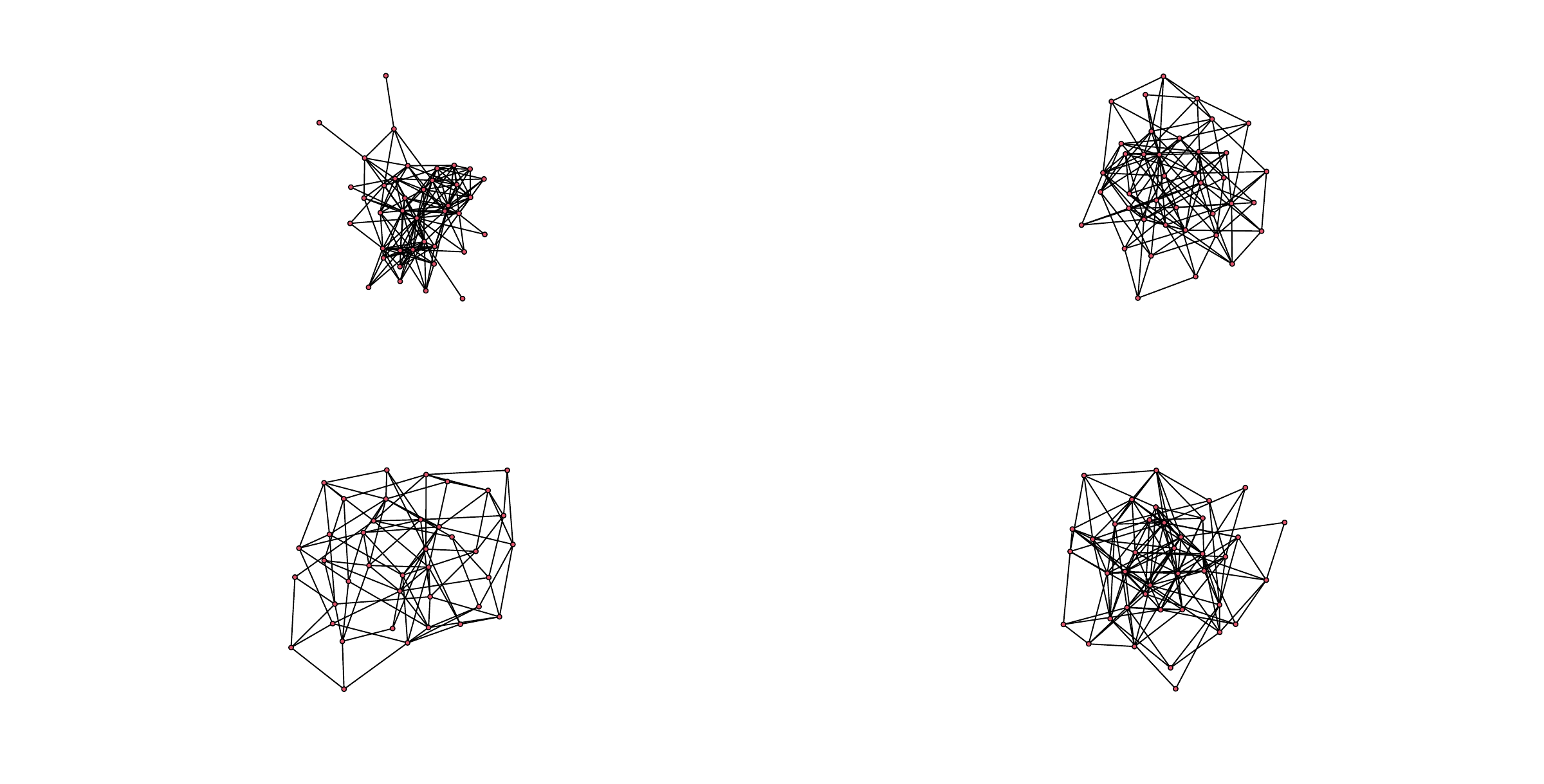}

    \caption{ Upper left network denotes the observed Kapferer network of tailor shop workers. The remaining three networks are draws from the ERGM model with equation $  P_{\boldsymbol{\theta}}(Y_{i,j}=1) = \psi({\boldsymbol{\theta}}) exp \{ \theta_0 \times edges + \theta_1 nsp(2) + \theta_2 esp(4) \}$. }
\end{figure}

\begin{figure}[ht]
    \centering

\includegraphics[width=15cm]{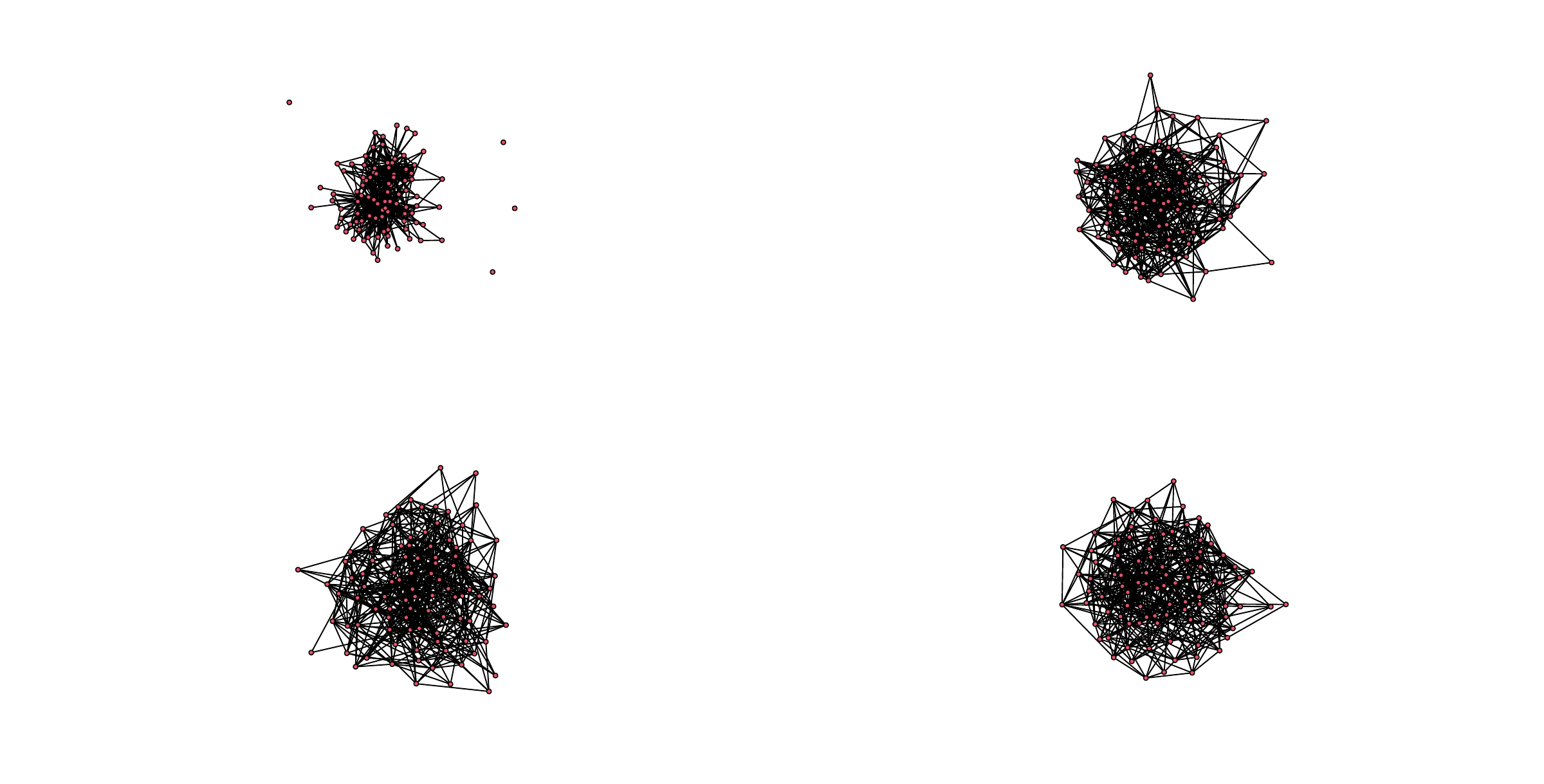}

    \caption{ Upper left network denotes the observed wind surfer network. The remaining three networks are draws from the ERGM model with equation $      P_{\boldsymbol{\theta}}(Y_{i,j}=1) = \psi({\boldsymbol{\theta}}) exp \{ \theta_0 edges + \theta_1 esp(18) + \theta_2 gwdsp \}$.  }
\end{figure}



\begin{figure}[ht]
    \centering

\includegraphics[width=15cm]{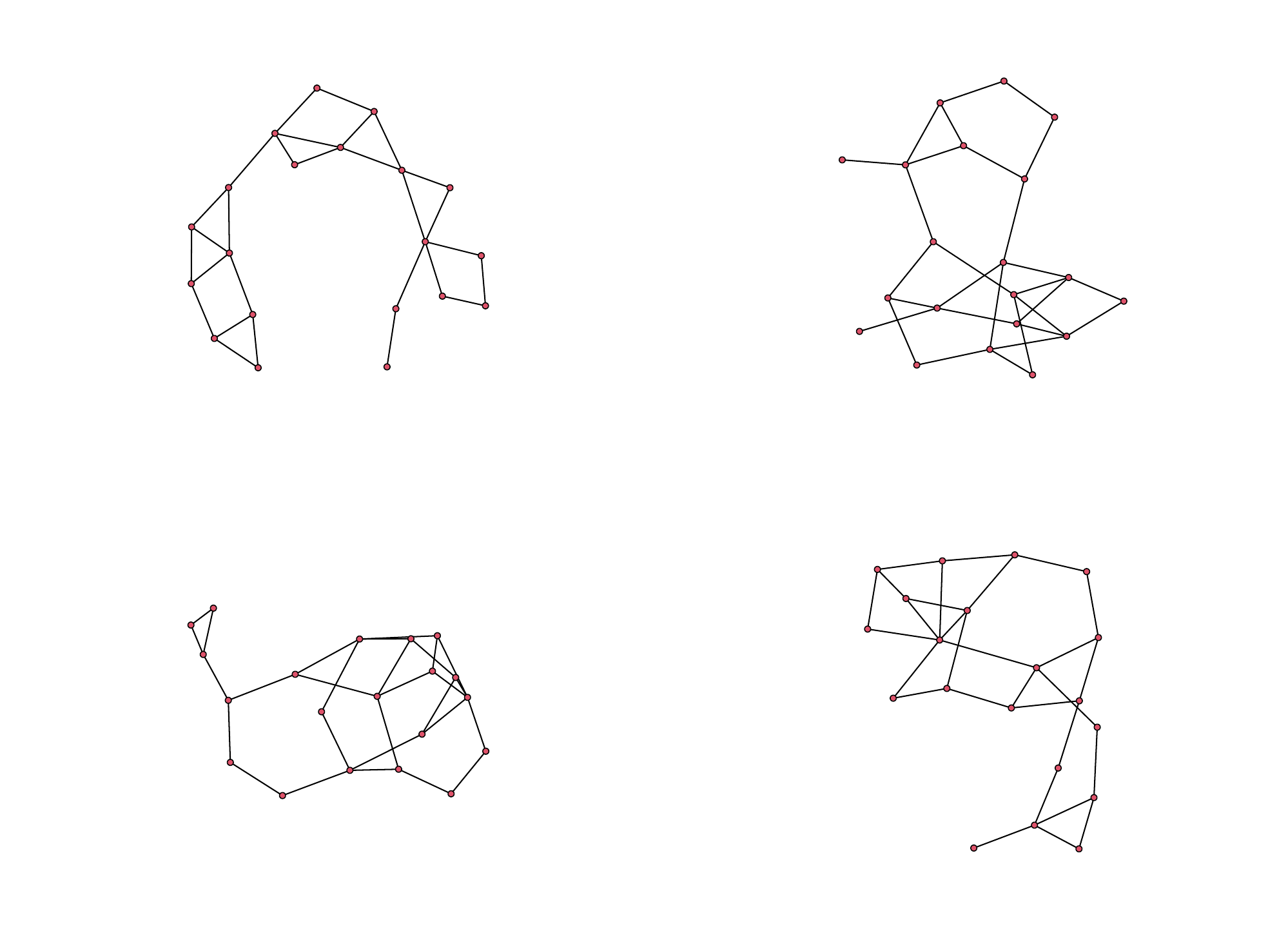}

    \caption{ Upper left network denotes the observed molecule network. The remaining three networks are draws from the ERGM model with equation $    P_{\boldsymbol{\theta}}(Y_{i,j}=1) = \psi({\boldsymbol{\theta}}) exp \{ \theta_0 edges + \theta_1 kstar(2) + \theta_2 kstar(4) \}$. }
\end{figure}


\begin{figure}[ht]
    \centering

\includegraphics[width=15cm]{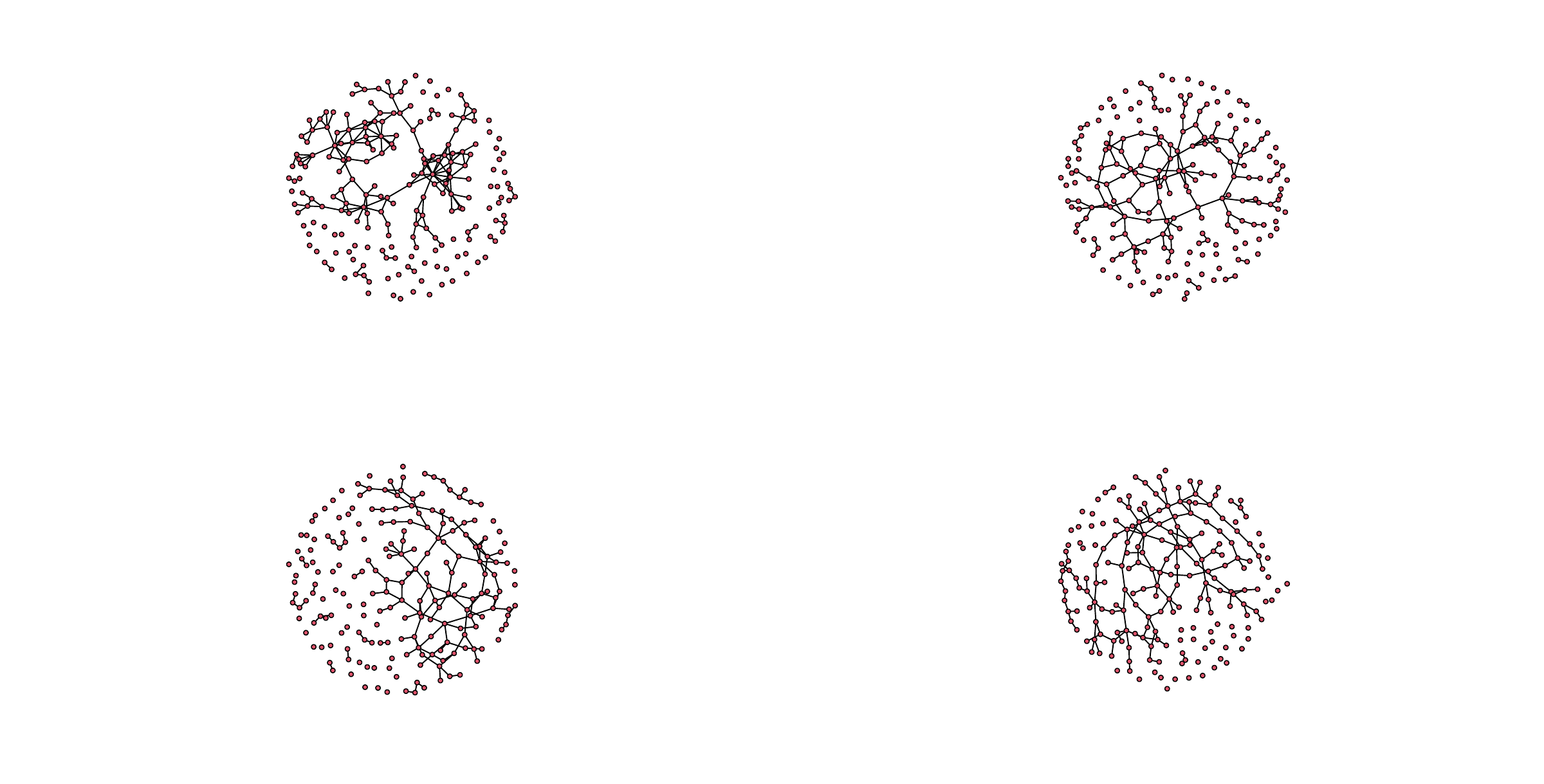}

    \caption{ Upper left network denotes the observed Faux Mesa high school network. The remaining three networks are draws from the  ERGM model with equation $        P_{\boldsymbol{\theta}}(Y_{i,j}=1) = \psi({\boldsymbol{\theta}}) exp \{ \theta_0 \times edges + \theta_1 nsp(4) + \theta_2 dsp(2) \}$.  }
\end{figure}

\begin{figure}[ht]
    \centering

\includegraphics[width=15cm]{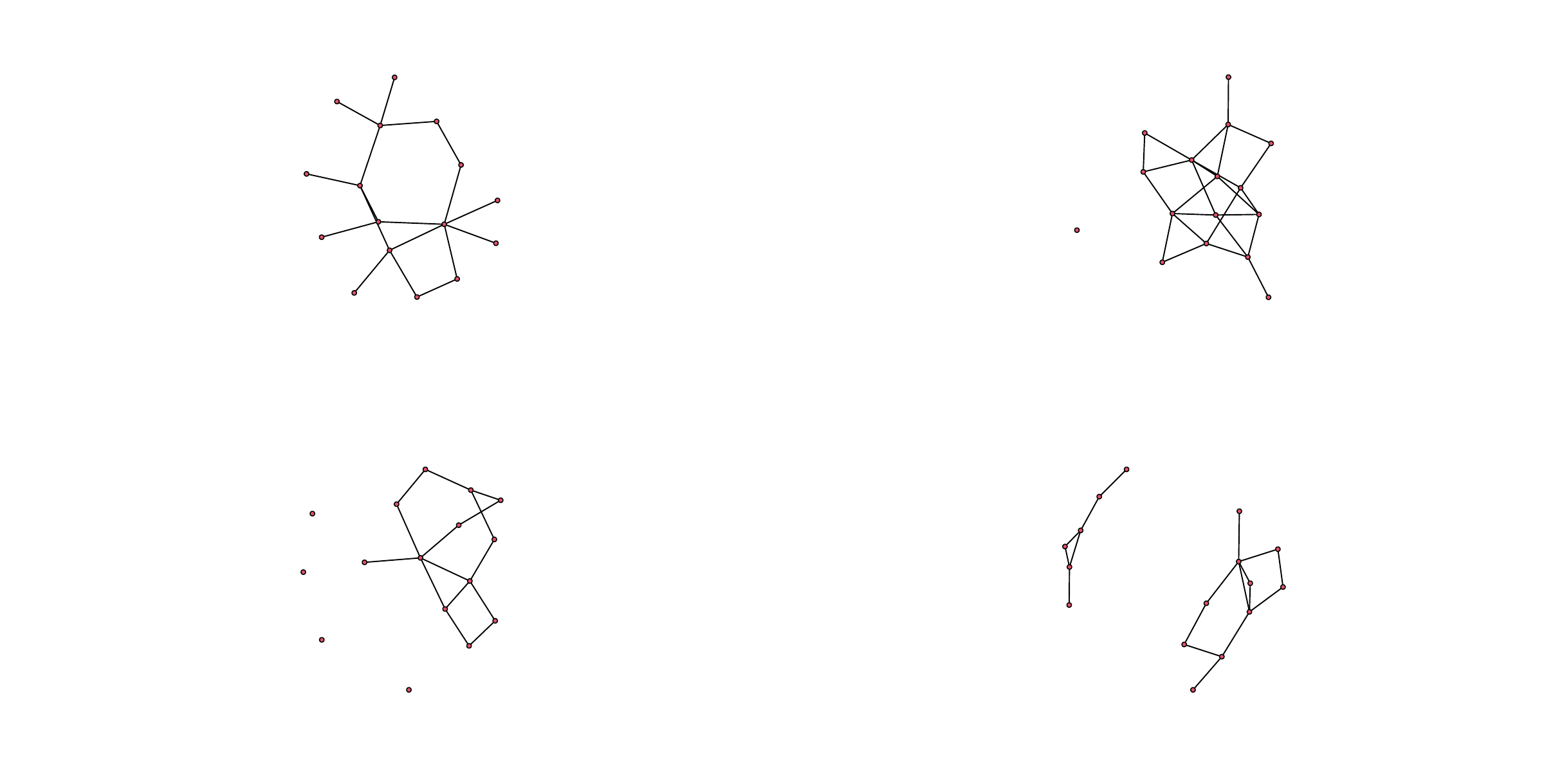}

    \caption{ Upper left network denotes the observed Moody contact network. The remaining three networks are draws from the ERGM model with equation $        P_{\boldsymbol{\theta}}(Y_{i,j}=1) = \psi({\boldsymbol{\theta}}) exp \{ \theta_0 \times edges + \theta_1 nsp(2) + \theta_2 degreepopularity \}$.  }
\end{figure}

\begin{figure}[ht]
    \centering

\includegraphics[width=15cm]{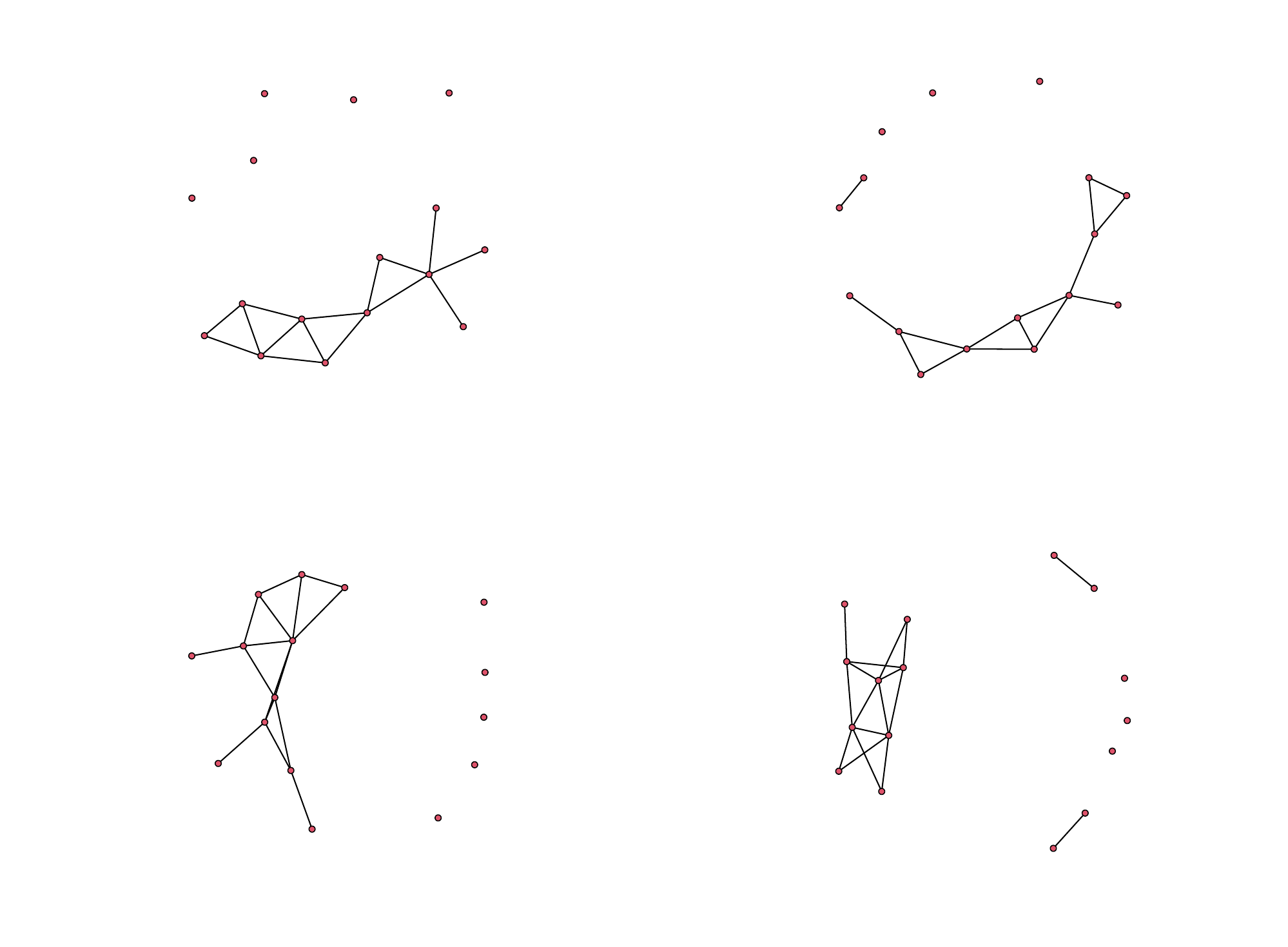}

    \caption{ Upper left network denotes the observed Florentine business network. The remaining three networks are draws from the  ERGM model with equation $    P_{\boldsymbol{\theta}}(Y_{i,j}=1) = \psi({\boldsymbol{\theta}}) exp \{ \theta_0 \times edges + \theta_1 isolates + \theta_2 gwesp \}$.  }
\end{figure}

\begin{figure}[ht]
    \centering

\includegraphics[width=15cm]{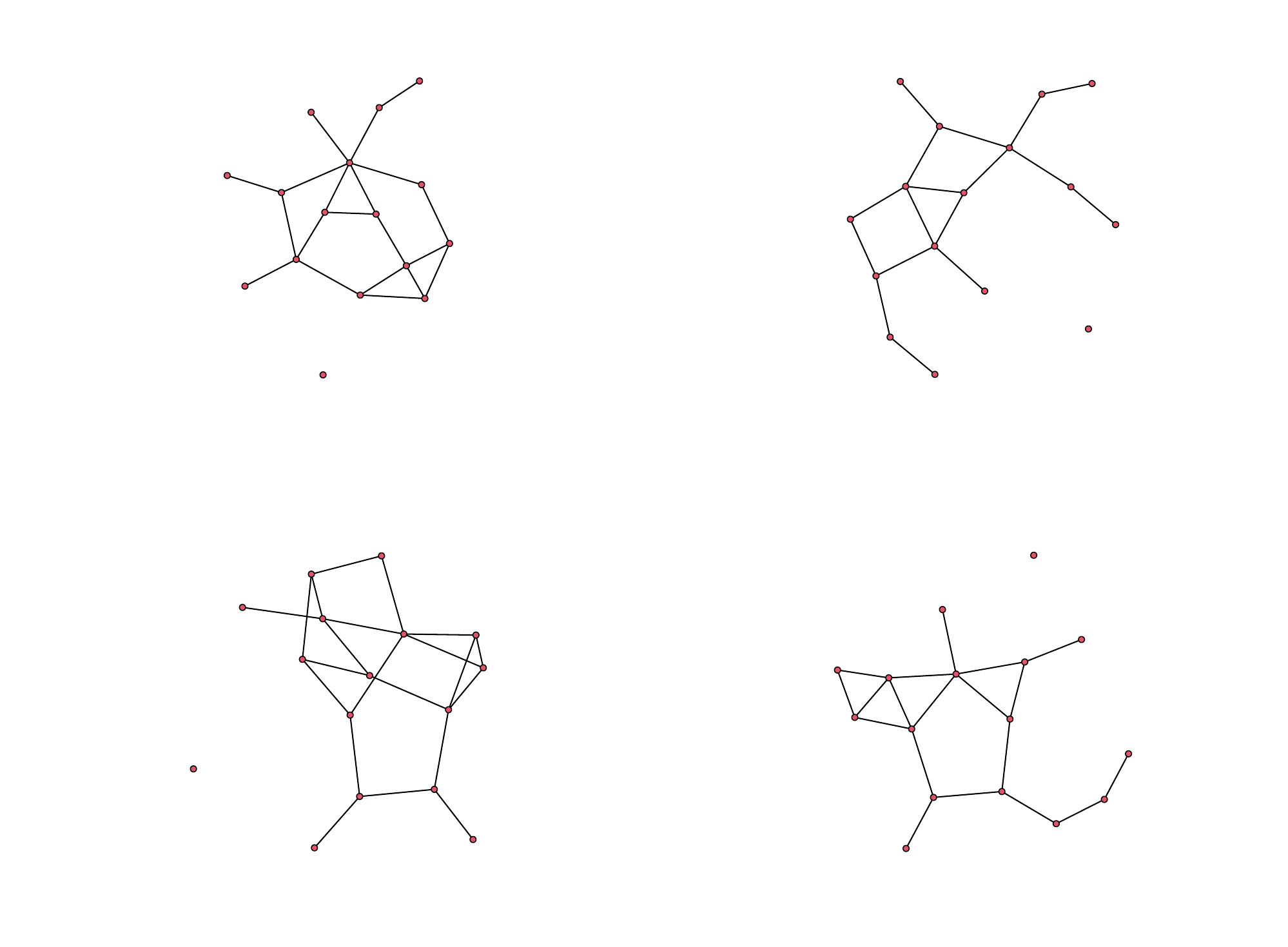}

    \caption{ Upper left network denotes the observed Florentine marriage network. The remaining three networks are draws from the  ERGM model with equation $    P_{\boldsymbol{\theta}}(Y_{i,j}=1) = \psi({\boldsymbol{\theta}}) exp \{ \theta_0 \times edges + \theta_1 g_1(x) + \theta_2 g_2(x) + \theta_3 g_3(x) \}$. $g_1(x)$ denotes the exogenous variable of familial wealth, $g_2(x)$ denotes the number of seats on the civic council and $g_3(x)$ denotes the total number of business and marriage ties.  }
\end{figure}


\end{document}